\documentclass[aps,prb,twocolumn,superscriptaddress,amsmath,eqsecnum]{revtex4}

\usepackage{dcolumn}
\usepackage{bm}
\usepackage{amsfonts}
\usepackage{graphicx,graphics}
\usepackage{rotate}
\usepackage{subfigure}
\usepackage{epsf}
\usepackage{epsfig}
\usepackage{color}
\usepackage{dcolumn}
\usepackage{color}
\usepackage{amsmath}
\usepackage{amssymb}

\newcommand{\f}[2]{{\ensuremath{\mathchoice%
        {\dfrac{#1}{#2}}
        {\dfrac{#1}{#2}}
        {\frac{#1}{#2}}
        {\frac{#1}{#2}}
        }}}
\newcommand{\ie}{{\it i.e. }}
\newcommand{\cf}{{\it cf.~}}

\newcommand{\eqn}[1]{\vspace{-0.cm}\begin{equation}
#1
\end{equation}}

\renewcommand{\=}{\,=\,}
\newcommand{\+}{\,+\,}
\renewcommand{\-}{\,-\,}

\newcommand{\moy}[1]{\ensuremath{\langle #1 \rangle}}

\begin{document}



\title{\textbf{Site Dilution in the Half-Filled One-Band Hubbard Model:
Ring Exchange,  Charge Fluctuations and Application to
La$_2$Cu$_{1-x}$(Mg/Zn)$_x$O$_4$
}}

\author{J.-Y. P. Delannoy}
\affiliation{Laboratoire de Physique, \'Ecole normale
sup\'erieure de Lyon, 46 All\'ee d'Italie, 69364 Lyon cedex 07,
France.}
\affiliation{Department of Physics and Astronomy, University of Waterloo, Ontario, N2L 3G1,
Canada}

\author{A.G. Del Maestro}
\affiliation{Department of Physics, Harvard University, Cambridge, Massachusetts, 02138, USA}

\author{M. J. P. Gingras}
\affiliation{Department of Physics and Astronomy, University of Waterloo, Ontario, N2L 3G1,
Canada}
\affiliation{Canadian Institute for Advanced Research, 180 Dundas Street
West, Toronto, Ontario, M5G 1Z8, Canada}
\affiliation{Department of Physics and Astronomy, University of Canterbury,
Private Bag 4800, Christchurch, New Zealand}

\author{P. C. W. Holdsworth}
\affiliation{Laboratoire de Physique, \'Ecole normale
sup\'erieure de Lyon, 46 All\'ee d'Italie, 69364 Lyon cedex 07,
France.}

\date{\today}

\begin{abstract}

We study the ground state quantum spin fluctuations around the N\'eel ordered
state for the one-band ($t,U$) Hubbard model on a site-diluted
square lattice. An effective spin Hamiltonian, $H_{\rm s}^{(4)}$, is
generated using the canonical transformation method, expanding to
order $t(t/U)^3$. $H_{\rm s}^{(4)}$ contains four-spin ring exchange
terms as well as second and third neighbor bilinear spin-spin  interactions.
Transverse spin fluctuations are calculated to order $1/S$ using a
numerical real space algorithm first introduced by Walker and
Walsteadt. Additional quantum charge fluctuations appear to this order in
$t/U$, coming from electronic hopping and
virtual excitations to doubly occupied sites. The ground state
staggered magnetization on the percolating cluster
decreases with site dilution $x$, vanishing 
very close to the percolation threshold. We compare our
results in the
Heisenberg limit, $t/U \rightarrow 0$, with quantum Monte
Carlo (QMC) results on the same model 
and confirm the existence of a systematic $x$-dependent difference
between $1/S$ and QMC results 
away from $x=0$. For finite $t/U$, we show that the
effects of both the ring exchange and charge fluctuations
die away rapidly with
increasing $t/U$. We use our finite $t/U$ results
to make a
comparison with results from experiments 
on La$_2$Cu$_{1-x}$(Mg/Zn)$_x$O$_4$.

\end{abstract}

\maketitle

\section{Introduction}

\subsection{Random disordered magnets}

Magnetic materials and model magnetic systems are perhaps the best
test benches for the study of collective phenomena in nature. This
is particularly true in the context of systems with frozen or
quenched random
disorder~\cite{Random-magnets,Ill-condensed-Matter}. Here,
questions such as the sharpness of phase transitions in disordered
systems~\cite{Harris-criterion}, the stability of ground-state
symmetry-breaking (random-field) perturbations~\cite{Imry-Ma} and
spin glass behavior arising from random
frustration~\cite{Binder-Young} have come under sharp scrutiny
over the past thirty years.

The 1987 discovery of high-temperature superconductivity in
doped antiferromagnetic copper oxide materials generated a
huge amount of interest in quantum
antiferromagnets which continues to this 
day~\cite{Anderson3,Sachdev-nature}. Here, the magnetic
properties depend strongly on the different possible types of
quenched disorder and this has proven to be a rich source of novel
quantum phenomena. An important area of investigation has been to
explore how the ground state of insulating quantum magnets evolves
as the level of random disorder is changed.  The following
examples represent a small subset of this class of studies.  A
large effort has been targeted towards understanding the
properties of antiferromagnetic spins chains subject to various
types of disorder~\cite{1d-chains,Eggert,Sirker}. Further work investigated how
long range order develops in two and three dimensional arrays of
weakly coupled integer (Haldane) spin chains~\cite{1d-chains} and
even-leg ladders~\cite{Azuma}. The question of how N\'eel order
develops upon magnetically diluting pure systems with quantum spin
liquid ground states is another field of intensive
 study~\cite{emerge-Neel}.

Theoretical problems relating to various types of random bond
disorder, as opposed to the more material-relevant case of site
dilution, have also been investigated~\cite{Vekic,Haas}. In three
dimensional systems, one noteworthy example is the so-called
antiglass phenomenon in LiHo$_x$Y$_{1-x}$F$_4$ where, for a low
concentration, $x$, of magnetic Ho$^{3+}$ ions, the dipolar
spin glass phase seemingly disappears~\cite{Ghosh}.
Another interesting problem concerns the role  frozen
random impurities may play at
conventional and deconfined quantum critical points in
two-dimensional antiferromagnets~\cite{Metlitski}.

However, among the multitude of interesting problems, a particular one,
possibly because of its seemingly simple physical setting and its
broad conceptual appeal, has drawn considerable attention: that of
the evolution of the antiferromagnetic N\'eel ground state in the
site-diluted $S=1/2$ nearest neighbor square lattice Heisenberg
 antiferromagnet (SLHAF).

\subsection{Site-diluted SLHAF and La$_2$Cu$_{1-x}$(Mg/Zn)$_x$O$_4$ }

As the insulating and antiferromagnetic parent
of high-temperature superconductivity in
La$_{2-x}$Sr$_x$CuO$_4$, 
La$_2$CuO$_4$ has quasi two dimensional magnetic 
exchange interactions and 
a good starting point for its description is to treat 
the CuO$_2$ planes as decoupled SLHAFs. 
Hence, experimental studies on zinc and magnesium
substitution for copper in
La$_2$CuO$_4$~\cite{Chakraborty,Cheong}, provided some of the
earliest motivation and interest in the problem of site-diluted
SLHAFs~\cite{Wan}. In particular, 
Cheong {\it et al.}~\cite{Cheong} found from bulk thermodynamic measurements
that in the diluted $S=1/2$ quantum antiferromagnetic materials,
La$_2$Cu$_{1-x}$Zn$_x$O$_4$ and La$_2$Cu$_{1-x}$Mg$_x$O$_4$, the
N\'eel temperature, $T_{\rm N}$, 
vanishes faster than in other
materials that can be considered as site-diluted classical  square lattice
magnetic systems
(either because they have large spin $S$, or because they have
large Ising anisotropies).

Most importantly, these early experimental results suggested 
that $T_{\rm N}$,
hence long range antiferromagnetic N\'eel order,
 may vanish at a critical impurity
concentration $x_c$ less than the site dilution percolation
threshold for the square lattice, $x_p\approx 0.41$. This
possibility was seemingly
supported by subsequent muon spin relaxation
($\mu$SR) and nuclear quadrupole resonance (NQR)
experiments~\cite{Corti}, with these latter measurements also
suggesting the possibility of a second transition below  $T_{\rm
N}(x)$ into a spin-glass like state.

From a
classical point of view, the ground state  of the SLHAF has two-sublattice N\'eel
order for all $x<x_p$. Consequently, 
early experiments on site-diluted La$_2$CuO$_4$ from Refs.~[\onlinecite{Cheong,Corti}]
 implied that either a
novel quantum ground state develops in the site-diluted SLHAF for $x_c<x<x_p$, or that
{\it frustrating} further neighbor exchange interactions
are important in the real material and that these
drive the
system into a two dimensional Heisenberg spin glass ground state,
presumably via the proliferation of Villain canted states for
$x_c<x<x_p$~\cite{Villain-ZPhysB,Kinzel,Saslow1,Saslow2,Vannimenus,Gawiec1}.

The idea that N\'eel order could disappear
in the diluted SLHAF, due to
quantum effects,
for a concentration of magnetic moments less than
the geometric site percolation threshold
($x<x_p$)
had  been suggested by some~\cite{Cherny-PRB}, but
not all~\cite{Wan}, early calculations. 
However, in strong
contrast to the early  body of experimental evidence~\cite{Cheong,Corti}
and theoretical suggestions~\cite{Cherny-PRB}, 
large scale quantum
Monte Carlo (QMC) simulations on the diluted SLHAF find that N\'eel
order survives up to the percolation threshold
$x_p$~\cite{Kato,Sandvik}. Further, contrary  to earlier
experiments~\cite{Cheong,Corti}, recent neutron scattering studies
on single crystals of La$_2$Cu$_{1-x}$(Mg/Zn)$_x$O$_4$ found that
long-range N\'eel order does survive up to at least $x=0.39$, if
not up to $x_p$ ~\cite{Vajk,Vajk-review}.
Interestingly, recent QMC studies show that the same scenario holds
for homogeneous bond dilution, with exotic quantum phases appearing only
for inhomogeneous dilution where local ladder structures form~\cite{Haas}.

A proposed explanation for the discrepancy between the earlier
experiments~\cite{Chakraborty,Cheong,Corti} and the more recent
ones~\cite{Vajk,Vajk-review} is that samples are extremely
sensitive to excess oxygen,  or off-stoichiometric
$\delta$, La$_2$CuO$_{4+\delta}$, as Cu$^{2+}$ is substituted by either
Zn$^{2+}$ or Mg$^{2+}$. Off-stoichiometry with $\delta>0$ is  
hole-doping, which is extremely detrimental to long range N\'eel order.
Thus, the present picture, supported by both
numerical~\cite{Wan,Kato,Sandvik} and
experimental~\cite{Vajk,Vajk-review} studies, is that N\'eel order
survives in the site-diluted SLHAF~\cite{Kato,Sandvik} and
La$_2$Cu$_{1-x}$(Mg/Zn)$_x$O$_4$ \cite{Vajk,Vajk-review} up to
$x_p$, with no intervening exotic quantum ground state for
$x<x_p$.

\subsection{Quantum Monte Carlo vs La$_2$Cu$_{1-x}$(Mg/Zn)$_x$O$_4$ }

\begin{figure}[ht]
\includegraphics[scale=0.7]{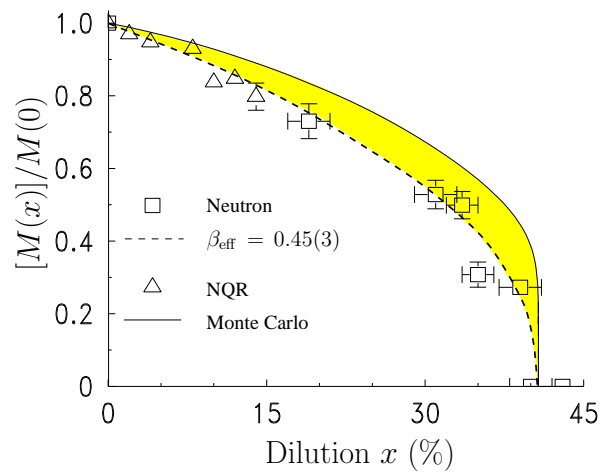}
\caption{Ground state staggered magnetization, $[M(x)]$, 
as a function of concentration, $x$,
of Zn and Mg in La$_2$Cu$_{1-x}$(Mg/Zn)$_x$O$_4$, normalized to the value for
zero dilution, $M(0)$ ~\cite{Vajk}. 
The solid line shows the results from quantum Monte
Carlo~\cite{Sandvik} for the site-diluted SLHAF. The figure is
reproduced from Ref.~[\onlinecite{Vajk}].}\label{science}
\end{figure}

While both high precision quantum Monte Carlo (QMC) studies of the
site-diluted SLHAF~\cite{Sandvik} and neutron scattering
experiments~\cite{Vajk,Vajk-review}
 on La$_2$Cu$_{1-x}$(Mg/Zn)$_x$O$_4$ now find that the
N\'eel order survives up to $x_c\approx x_p$ (exactly $x_c=x_p$
for the QMC simulations), the quantitative agreement stops here.
There is a systematic discrepancy between QMC and the neutron
results for the sublattice N\'eel order parameter, $[M(x)]$,
as a function of $x$. The experimental and numerical data are reproduced in
Fig.~\ref{science}.  In this figure, the QMC results of
Ref.~[\onlinecite{Sandvik}] are shown by the upper solid line. The
experimental results (neutron, squares, from Ref.~[\onlinecite{Vajk}]; NQR,
triangle, from Ref.~[\onlinecite{Corti}])
 lie on the dashed line, which is
a guide to the eye parameterized by 
$[M(x)]/M(0)  = (1-x/x_p)^{\beta_{\rm eff}}$. The QMC results lie
above the experimental data over the whole range $0<x<x_p$, as
illustrated by the shaded region. 
Taking it as a premise that the
QMC data are essentially the exact results for the diluted $S=1/2$
SLHAF, the systematic difference between them and the experimental
data shown in Fig.~\ref{science} suggests that Zn$^{2+}$ and
Mg$^{2+}$ substituted Cu$^{2+}$ in La$_2$CuO$_4$ are not
quantitatively described by a site-diluted nearest neighbor Heisenberg
Hamiltonian. 
The nature of the discrepancy is in itself
interesting. It is initially small at low $x$,  increases and
reaches a maximum for $x\sim 0.35$, and decreases upon approaching
$x_p$ such that the ``true'' underlying microscopic Hamiltonian describing
La$_2$Cu$_{1-x}$(Mg/Zn)$_x$O$_4$ seem to also 
possess a percolation threshold
very close to that of the idealized SLHAF.

\subsection{Ring exchange interactions}

One class of candidate  perturbations that may be giving the missing physics
of diluted La$_2$CuO$_4$ are ring, or cyclic, exchange interactions
involving multiple interactions around closed plaquettes of the
square lattice. Such interactions 
have received intensive attention
 recently~\cite{Notbohm,Gossling,Roger-proc} and 
have been shown to play an
important role in the quantitative description of undiluted 
La$_2$CuO$_4$~\cite{coldea01,Toader,AMT_comment_Toader,Toader-reply}.
%
%
Taking as a starting point the one-band half-filled Hubbard
model~\cite{mac,trem,Delannoy}, the lowest order ring exchange interaction
takes its origin in  
virtual electronic hopping process, fourth order in $t/U$, taking
electrons coherently around a closed square plaquette. 
Here $t$ is the nearest neighbor hopping constant and
 $U$ is the on-site Coulomb energy.
Taking it as plausible~\cite{Toader,AMT_comment_Toader,Toader-reply}
 that ring exchange is indeed present
and a leading perturbation beyond the Heisenberg model
description of La$_2$CuO$_4$, it is natural to ask
what  its effect is on the N\'eel order parameter upon
substituting Cu by a 
concentration $x$ of non-magnetic ions (see Fig.~\ref{science}).
This question, which to the best of our knowledge has so far not been
investigated, is the one that we explore in this paper.

To tackle this question, one must return to a problem of correlated
electrons. 
The reason is that the spin-only Hamiltonian with ring exchange derives
from a set of electronic hops. As we show below,
the elimination of an intermediate 
site in an electron hopping pathway, affects
the resulting effective spin Hamiltonian in a nontrivial manner.
Specifically, we consider the problem of a site-diluted
half-filled one-band Hubbard model away from the Heisenberg
$t/U\rightarrow 0$ limit.
Since here ring exchange originates solely 
from correlated nearest
neighbor electronic hops, they cannot move the percolation
threshold to a larger value than the nearest neighbor threshold
$x_p$. From this constraint alone, ring exchange is an admissible
candidate for a perturbation to the diluted $S=1/2$ SLHAF,
as it preserves the same geometric percolation threshold $x_p$ as
the nearest neighbor Heisenberg model.

%

The presence of the ring exchange and  second 
and third nearest neighbor bilinear exchange terms in the Hamiltonian, generated by hopping
processes to fourth order in $t/U$, leads to a sign problem for
currently available 
QMC methods  using
the standard $S^z$ basis representation of the Hamiltonian~\cite{Melko-Kaul}.
A direct attack on the site-diluted ring exchange Hamiltonian via
QMC, such as done for the site-diluted Heisenberg model~\cite{Sandvik}
is therefore not possible at this time.
As a first step in investigating the role played by ring exchange
in the site-diluted Hubbard model, 
we carry out a finite-lattice spin wave
calculation to order $1/S$ on an extended effective 
spin Hamiltonian generated from up to four hop
electronic pathways. To proceed, we use
a real space linear spin wave method adapted
to finite-size diluted lattices, first developed by Walker and
Walsteadt~\cite{ww} in the context of spin glasses
and similar to that used for the site-diluted
nearest neighbor Heisenberg antiferromagnet on the square~\cite{Castro} 
and honeycomb lattices~\cite{Beach}. We investigate
the role of ring exchange on the 
dependence of the ground state staggered magnetization,
$[M(x)]$, as a function of $x$. 
In Ref.~[\onlinecite{Castro}], it was found
that there is a systematic difference 
between the value of this quantity, calculated via
the spin wave method and the essentially exact QMC~\cite{Sandvik}.
From this, it is clear that a similar systematic difference should
also exist between our data 
calculated using a $1/S$ expansion, and 
what would be the not yet available numerically
exact value for $[M(x)]$, as a
function of dilution, for the extended Hamiltonian. 
Hence, although  the main motivation for this project comes from the experiment
on  La$_2$Cu$_{1-x}$(Mg/Zn)$_x$O$_4$ ~\cite{Vajk},
some care has to be taken in attempting to make
a direct comparison  with experimental  results. 
Rather, our results for
the extended Hamiltonian and electronic hopping, can
be quantitatively benchmarked by a comparison with those for the site-diluted 
Heisenberg model, using the same real space expansion technique.
From a broader perspective, our work provides a first
glimpse at the role  of charge correlation effects in the problem of 
diamagnetic site dilution in the one-band Hubbard model.

\subsection{Charge fluctuations}

The generation of ring and further neighbor 
exchange interactions is not the only effect
of extending the analysis of the one-band Hubbard 
model beyond the Heisenberg limit
using a perturbation expansion in $t/U$. 
We have previously shown that extending the expansion to 
order $(t/U)^4$ generates quantum charge 
fluctuations \cite{Delannoy} that are independent
of the transverse spin fluctuations of localized 
$S=1/2$ moments. These fluctuations appear
in the perturbation expansion on the square lattice 
because, to this order,
the ground state wave function contains 
an admixing with excited states 
corresponding to doubly occupied sites.
As doubly occupied sites carry no moment,
 the expectation value for the magnetic 
moment of the Hubbard model is reduced below
that expected from the effective spin-only
Hamiltonian describing transverse spin fluctuations. We show here 
that these charge fluctuations are a key 
element in the ultimate success of comparisons between the
one-band Hubbard model and experiments on both 
undiluted and site-diluted La$_2$CuO$_4$. 
Just as for ring exchange effects, 
we find that the effects of charge fluctuations disappear as the
site percolation threshold is approached, 
as four hop electronic processes are interrupted by the
dilution well before this limit is reached.

The rest of the paper is organized as follows:
before launching into the calculations, we discuss in Sections IIA
and IIB some of the caveats that arise when considering a low
energy effective spin-only Hamiltonian derived from a
site-diluted Hubbard model. In particular, the exchange
interactions become explicitly disorder dependent (Section IIA).
Furthermore, by going beyond the Heisenberg limit, the operator for the N\'eel
order parameter has to be corrected to take into account the
charge mobility of the electrons in the Hubbard
model~\cite{Delannoy}. 
The results presented below show that this correction is crucially 
important to obtain the correct $1/S$ behavior of the model.
The consequent reduction in the amplitude
of the staggered magnetization in the presence of local disorder is
discussed in Section IIB. We then discuss in Section IIC the
stability of the classical N\'eel ground state for finite
disorder, when ring exchange is present. Section IID describes the
spin wave method that we use. Section III gives an overview of the
algorithmic procedure used to diagonalize the quadratic form of
the disordered finite-lattice spin Hamiltonian. The numerical
results are presented in Section IV, followed in Section V 
by a discussion of the results and a perspective for future work.
An appendix discusses the question of statistical
uncertainties in the data presented in Section IV.

\section{Spin Hamiltonian and Real Space Linear Spin Wave Calculation}

\subsection{Spin Hamiltonian}
We begin with the Hubbard Hamiltonian, $H_{\rm H}$:

\begin{eqnarray}
H_{\rm H} &= &T + V \\ & = &  -t \sum'_{i,j;\sigma} \epsilon_i
\epsilon_j c^\dagger_{i,\sigma}c^{\phantom{\dagger}}_{j,\sigma} \+ U \sum'_i
\epsilon_i n_{i,\uparrow}n_{i,\downarrow} . \label{HH}
\end{eqnarray}

The first term is the kinetic energy term that destroys an
electron of spin $\sigma$ at site $j$ and creates one on the
nearest neighbor site $i$. The second term is the on-site Coulomb
energy $U$ for two electrons with opposite spins to be on the same
site $i$ and where $n_{i,\sigma}=c^\dagger_{i,\sigma}c_{i,\sigma}$
is the occupation operator at site $i$.  A site $i$ substituted by
a non-magnetic cation has $\epsilon_i=0$, otherwise
$\epsilon_i=1$. In the following we use the notation $\sum'$ to
represent a summation over the $L^2$ sites of the square lattice
and $\sum$ for a sum over the $N=\sum'_i (1-\epsilon_i)$ undiluted
sites. The number of magnetic sites and hence of mobile electrons, $N$,
at half filling, is thus configuration dependent. The average
concentration of vacancies is $1-[\epsilon_i]_{\rm disorder}=x$.
Similarly, $\sum'_j$ represents a sum over neighboring sites and
$\sum_j$ a sum over neighboring occupied magnetic sites. Below, a summation index with
angular brackets; $\langle \dots \rangle$ in $\sum_{ \langle \ldots \rangle }$
denotes an ordered sum, taking into account only unique pathways.

The derivation of a spin Hamiltonian from a one-band Hubbard model
can be performed through many different methods, leading to
apparently different effective spin Hamiltonians. It is only
recently that it has been shown \cite{trem} that all these
Hamiltonians are equivalent, as they are related to each other
through a unitary transformation. We have recently applied the
canonical transformation method, which
uses the ratio $t/U$ as a small parameter in a perturbation
expansion, to study the magnetic excitations and the
staggered magnetization in the Hubbard model~\cite{Delannoy,Delannoy2}. 
The method, introduced by Harris 
{\it et al.}
\cite{Harris} and developed further by MacDonald {\it et al.}
\cite{mac,mac2,mac3}, relies on the separation of the kinetic part $T$ of
the Hubbard Hamiltonian into three terms that respectively
increase by one ($T_1$), keep constant($T_0$) or decrease by one
($T_{-1}$) the number of doubly occupied sites.
Specifically, one writes:
%
\eqn{T\,=\,-t\sum'_{i,j;\sigma} \epsilon_i\epsilon_j
c^\dagger_{i,\sigma}c^{\phantom{\dagger}}_{j,\sigma}\,=\,T_1\,+\,T_0\,+\,T_{-1}}
\begin{eqnarray}
T_1 &= &-t \sum'_{i,j;\epsilon_i\epsilon_j 	\sigma}n_{i,\bar{\sigma}}
\epsilon_i\epsilon_j  c^\dagger_{i,\sigma}c^{\phantom{\dagger}}_{j,\sigma} h_{j,\bar{\sigma}}\\
T_{0}&=&-t \sum'_{i,j;\sigma}	\Big( \epsilon_i\epsilon_j  h_{i,\bar{\sigma}}
c^\dagger_{i,\sigma}c^{\phantom{\dagger}}_{j,\sigma}
h_{j,\bar{\sigma}} \nonumber \\&&\qquad +\, n_{i,\bar{\sigma}}
c^\dagger_{i,\sigma}c^{\phantom{\dagger}}_{j,\sigma} n_{j,\bar{\sigma}} \Big) \\
T_{-1}&=&-t \sum'_{i,j;\sigma} \epsilon_i\epsilon_j  h_{i,\bar{\sigma}}
 c^\dagger_{i,\sigma}c^{\phantom{\dagger}}_{j,\sigma}
n_{j,\bar{\sigma}}
\end{eqnarray}
where $\bar{\sigma}$ stands for up if ${\sigma}$ is down and for
down if $\sigma$ is up and where $h_{i,\overline{\sigma}} = 1-n_{i,\overline{\sigma}}$. This separation comes from multiplying the
kinetic term  $T$ on the right by $n_{i,\bar{\sigma}} +
h_{i,\bar{\sigma}} \,=\,1$ and multiplying on the left by
$n_{j,\bar{\sigma}} + h_{j,\bar{\sigma}} \,=\,1$.

\begin{figure}[ht]
\includegraphics[scale = 1]{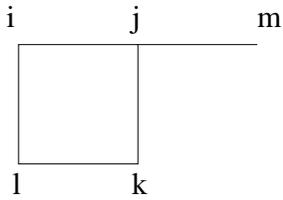}
\caption{Labels for the different sites involved in the effective
spin interactions and arising from a $t-U$ Hubbard model up to
order $t(t/U)^3$. 
$\langle i,j \rangle$, $\langle \langle  i,k \rangle \rangle$, $\langle \langle \langle  i,m \rangle
\rangle \rangle$ are nearest,
second nearest and third nearest neighbors, respectively.
$\langle i,j,k,l \rangle$ denotes the sites that belong to an elementary square
plaquette. } \label{fig2}
\end{figure}

Applying a unitary transformation ${\rm e}^{i{\cal S}}$ to $H_{\rm H}$
leads to a spin-only Hamiltonian through the relation:
\eqn{H_{\rm s}\,=\,{\rm e}^{i{\cal S}} H_{\rm H} {\rm e}^{-i{\cal
S}}\,=\, H_{\rm H} + \f{[i{\cal S},H_{\rm H}]}{1!} +  \f{[i{\cal S},[i{\cal
S},H_{\rm H}]]}{2!} + \cdots  , \label{commut}}
We do not reproduce the derivation here, rather we refer the
reader to Refs.~[\onlinecite{mac,Delannoy,Delannoy2}] for details of
the form of $\cal S$ and $H_{\rm H}$ order by order in the development.
Up to third order in the $t/U$ expansion, 
we finally find for the effective spin Hamiltonian:
\begin{eqnarray}
H_{\rm s}^{(4)} & = &  \sum_{\langle i,j \rangle}J_1\left(\bm{S}_i  \cdot \bm{S}_j \right) \nonumber\\
& + & \sum_{\langle \langle i,k \rangle \rangle }J_2\left(\bm{S}_i  \cdot \bm{S}_k \right) \nonumber\\
& + & \sum_{\langle \langle  \langle i,m \rangle \rangle \rangle }J_3\left(\bm{S}_i  \cdot \bm{S}_m \right)\label{SS} \\
& + & \sum_{\langle i,j,k,l\rangle}J_c\left\{(\bm{S}_i \,\cdot\,\bm{S}_j)\,(\bm{S}_k\,\cdot\,\bm{S}_l)\right.\nonumber\\
&&\left.
+(\bm{S}_i\,\cdot\,\bm{S}_l)\,(\bm{S}_k\,\cdot\,\bm{S}_j)\,
-\,(\bm{S}_i\,\cdot\,\bm{S}_k)\,(\bm{S}_j\,\cdot\,\bm{S}_l)\right\},
\nonumber
\label{Hs4}
\end{eqnarray}
where the site labels refer to the configuration shown in Fig.~\ref{fig2}
The different coupling constants $(J_1,J_2,J_3,J_c)$ arise as a
result of the integration over all electronic paths allowed in the
site-diluted Hubbard model. As a result, they depend on the local
site occupancy along the exchange path. We find:
%
%
%
\begin{equation}
J_1  = 4 \left\lbrack\f{t^2}{U} \epsilon_i \epsilon_j \,-\,
\f{t^4}{U^3} \left( 4 \epsilon_i \epsilon_j + {\cal N}_{ij}
\right)\right\rbrack,
\label{J1}
\end{equation}
\begin{equation}
J_2  =  4 \left\lbrack \f{t^4}{U^3} \left( \epsilon_i \epsilon_j
\epsilon_k + \epsilon_i \epsilon_l \epsilon_k  -
 {\cal N}_{ik}
\right)\right\rbrack ,
\label{J2}
\end{equation}
\begin{equation}
J_3  = 4 \left\lbrack\f{t^4}{U^3}  \epsilon_i \epsilon_j
\epsilon_m \right\rbrack , 
\label{J3}
\end{equation}
\begin{equation}
 J_c = 80
\left\lbrack\f{t^4}{U^3} \epsilon_i \epsilon_j \epsilon_k \epsilon_l
\right\rbrack,
\label{Jc}
\end{equation}
where ${\cal N}_{\mu\nu}$ is a plaquette index for bond $\mu\nu$
and is equal to the number of plaquettes to which both sites $\mu$
and $\nu$ belong. When there is no dilution, ${\cal N}_{\mu\nu}=2$
for all  nearest neighbor $\langle i,j \rangle$ bonds and $1$ for second neighbor bonds
$\langle \langle i,j \rangle \rangle$ across the diagonal of a plaquette. When one of the four bonds
defining a plaquette is missing, the ${\cal N}_{\mu\nu}$ for the
three nearest neighbor bonds along the remaining edges of the
plaquette are reduced from two to one.  ${\cal N}_{\mu\nu}$ for
the next nearest neighbor bond across the diagonal of the
plaquette is reduced from one to zero. For example, consider 
Fig. \ref{fig2} where only the site $j$  has been eliminated 
by dilution. The expression for the coupling
constants becomes: 
\eqn{
\begin{array}{lcl}
J_1(i,l) & = & 4\f{t^2}{U}-20\f{t^4}{U^3} \\
\\
J_2(i,k) & = & 4 \f{t^4}{U^3} \\
\\
J_3(i,m) & = & 0 \,\,\,\,\,\,\,\,\,\,\,\,J_c(i,j,k,l) \=  0,
\end{array}
\label{Jij-dil}
}
which should be compared with $J_1=4t^2/U-24t^4/U^3$,
$J_2=J_3=4t^4/U^3$ and $J_c=80t^4/U^3$ for the undiluted lattice.
The most important point here is that since the antiferromagnetic
and frustrating $J_2$ and $J_3$ exist solely via electronic
hopping processes connecting nearest neighbor sites, these
interactions are progressively eliminated
as intermediate sites are diluted.  That is,
if both sites $j$ and $l$ are missing then $J_2(i,k)=0$. Hence,
one can see that site dilution strongly affects the coupling
constants as further neighbor exchange depends on the existence of
a nearest neighbor pathway between the sites. This would not be
the case if the original Hubbard model included direct second
or third 
nearest neighbor hopping
 parameters, $t'$ and $t''$, respectively~\cite{Delannoy2}.
We will return to this issue in the Conclusion section.
However, in this
paper we limit ourselves to nearest neighbor hopping only.

\subsection{N\'eel order parameter}

Our objective is to calculate the ground state N\'eel order
parameter for the original Hubbard model as a function of site
dilution, using a spin-only description. To do this, the
staggered (spin density wave) magnetization operator,
\eqn{\hat M_{\rm H}\,=\,\f{1}{N}\sum_{i} (-1)^i (n_i^{\uparrow}-n_i^{\downarrow})
,\label{Mag}}
%
defined for the Hubbard model, 
must be canonically transformed before it can be exploited
in a spin-only description.
Here,  $\hat M_{\rm H}$,
$\hat M_{\rm s}$  and
${\hat{\widetilde M}_{\rm s}}$ 
refer to operators while $M_{\rm s}$ and 
${{\widetilde M}_{\rm s}}$ refer to
their expectation values.
That is, within the effective theory $M_{\rm H}$ becomes
$\hat M_{\rm s}=e^{i {\cal S}} \hat M_{\rm H} e^{-i {\cal S}}$ and
the expectation value in the ground state is defined
\begin{equation}
M_{\rm s}\,=\, \frac{_{\rm H}\langle 0|\hat M_{\rm
H}|0 \rangle_{\rm H}} {_{\rm H}\langle 0|0 \rangle_{\rm H}}
\,=\,\frac{_{\rm s}\langle 0|{\hat M_{\rm s}}|0\rangle_{\rm s}}
{_{\rm s}\langle 0 |0\rangle_{\rm s}} \;\;\; . \label{O2}
\end{equation}
Here $|0\rangle_{\rm H}$ and $|0\rangle_{\rm s}=e^{i{\cal
S}}|0\rangle_{\rm H}$ are the ground state wave vectors in the
original Hubbard and spin-only models. We have recently
shown~\cite{Delannoy} that this is more than just an academic point. 
Rather, it has important consequences for the ground state
magnetization as one moves into the intermediate coupling regime and,
as we will show below, plays a significant quantitative role in
the present site-diluted Hubbard model.
As we apply the canonical transformation on 
$\hat M_{\rm H}$~\cite{Delannoy,Delannoy2} we find for $\hat M_{\rm s}$:
\eqn{\hat M_{\rm s}\,=\, \hat M_{\rm H} \+ \f{1}{U} \left(\tilde{T}_1
-\tilde{T}_{-1}\right)\+ \f{1}{2U^2}\left(\tilde{T}_{-1}T_1 -
T_{-1}\tilde{T}_{1}\right),\label{effmag}}
where
\begin{eqnarray}
\f{1}{N}\tilde{T}_1    & \equiv & [T_1,\hat M_{H}],\label{commut1}\\
\f{1}{N}\tilde{T}_{-1} & \equiv & [T_{-1},\hat M_{H}],\label{commut2}\\
\f{1}{N}\tilde{T}_0    & \equiv & [T_0,\hat M_{H}].\label{commut3}
\end{eqnarray}
After some algebra, we can
write this expression in terms of $S=1/2$ spin
operators\cite{Delannoy} as:
\begin{eqnarray}
\hat M_{\rm s} & = &  \f{1}{N}\sum'_i \epsilon_i S_i^z (-1)^i 
 \nonumber \\ &-&
 \f{2t^2}{N U^2} \sum'_{\langle i,j\rangle}\epsilon_i \epsilon_j \left\{ S_i^z \,-\, S_j^z\right\}
(-1)^i	 .  \label{one}
\end{eqnarray}

Recalling the standard definition for the staggered
magnetization operator in a spin model,
\begin{equation}
{\hat{\widetilde M}_{\rm s}}  =   \f{1}{N} \sum_i' \epsilon_iS_i^z (-1)^i ,
\label{one-tilde}
\end{equation}
%
%
we arrive at the principal result of Ref.~[\onlinecite{Delannoy}]
that
\begin{eqnarray}
M_{\rm s} & = & 
\frac{_{\rm s}\langle 0|{\hat M_{\rm s}}|0\rangle_{\rm s}}
{_{\rm s}\langle 0 |0\rangle_{\rm s}} \nonumber \\
{\widetilde M_{\rm s}}  & = &  
\,\frac{_{\rm s}\langle 0|{\hat {\widetilde M}_{\rm s}}|0\rangle_{\rm s}}
{_{\rm s}\langle 0 |0\rangle_{\rm s}} \nonumber \\
{\rm and} & & \nonumber \\ 
  M_{\rm s} & \ne & {\widetilde M_{\rm s}} .
\end{eqnarray}\label{unequal}
%
The difference is due to the appearance of new quantum fluctuations
arising from the charge delocalization over closed virtual loops
of electronic hops, which is the origin of the
second term in Eq.~(\ref{one}). These spin independent fluctuations, which
appear to order $t^2/U^2$ in the magnetization operator, are
generated when the canonical transformation is applied on the
Hamiltonian to order $t^4/U^3$ and are therefore not present in
the $(t/U\rightarrow 0)$ Heisenberg limit. 
We recently investigated the effects of these terms in the undiluted
case~\cite{Delannoy,Delannoy2}. Here, the disorder is manifest through the
dilution variables $\epsilon_i$ and, in this paper, we are
interested in how the spin renormalization factor modifies the
ground state magnetization upon site dilution. However, 
before doing so, we first return 
to a discussion of the ground state of 
the spin-only Hamiltonian $H_{\rm s}^{(4)}$.

Henceforth, for the sake of compactness, we shall omit the subscript ``s''
in $M_{\rm s}$ and ${\widetilde M}_{\rm s}$,
 understanding that all results presented below
were obtained from calculations performed on a spin-only 
description of the low-energy sector of the half-filled Hubbard model.

\subsection{Classical ground state}\label{pathways}

\subsubsection{$J_1$ interactions only}

\begin{figure}[ht]
\includegraphics[scale=0.7]{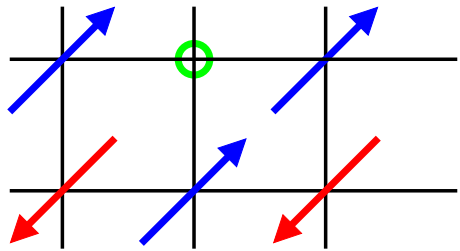}
\caption{Diluted N\'eel configuration. The circle labels a missing (diluted) site.
}\label{Neel_dilut}
\end{figure}

The real space spin wave method that we use to deal with dilution
requires, as the starting point, the knowledge of the
classical ground state spin
configuration. With nearest neighbor interactions only, the
classical ground state configuration is, in the absence of
dilution, the N\'eel staggered spin configuration. This long range ordered
state results from the local minimization of the
exchange interactions. Since we work with a concentration of
defects, or dilution, $x$, smaller than the  percolation
threshold $x_p$, there exists a percolating cluster of magnetic sites
with an exchange path connecting every pair of spins on the
cluster. As a result, the classical ground state configuration for the percolating
cluster is a connected N\'eel configuration, where every spin
keeps the orientation it would have had without dilution (see Fig.
\ref{Neel_dilut}).

\subsubsection{Full Hamiltonian}

In the case of the effective spin-only Hamiltonian, expressed in Eq. (\ref{SS}), the
situation becomes more complicated. If the $J_2$, $J_3$ or $J_c$
interactions get too large, the system undergoes a phase
transition to a new classical state that is not collinear.


\paragraph{Non diluted case\\}

As can be read from Eqs.~(\ref{J1},\ref{J2},\ref{Jc}),
when there is no dilution, the coupling constants read:
\eqn{\left\{\begin{array}{ccl}
J_1 & = & 4\f{t^2}{U} - 24 \f{t^4}{U^3}\\
J_2 & = & J_3 = \, 4 \f{t^4}{U^3} \\
J_c & = & 80 \f{t^4}{U^3}\\
\end{array}\right.
}
For $t/U\=1/8$, a value similar to that reported for La$_2$CuO$_4$
and that we henceforth take in the present work
\cite{coldea01,Delannoy2}, the ratios between the different
coupling constants are:
\eqn{\left\{\begin{array}{ccl}
\f{J_2}{J_1} & \simeq & 0.0172,\\
\f{J_3}{J_1} & \simeq & 0.0172,\\
\f{J_c}{J_1} & \simeq & 0.0862.\\
\end{array}\right.
}
For a model with nearest and next nearest couplings only, the
$J_1/J_2$ model, the N\'eel state is stable for
${J_2}/{J_1}\leqslant 0.5$ ~\cite{Chandra,Dagotto,houches}. For the
$J_1/J_c$ model, the quantity $\displaystyle \tilde{J_c}\,=\,J_c\,
S^2$ is usually introduced, and as long as $\displaystyle
\tilde{J_c}\,\leqslant\,{J}/{2}$, the N\'eel state is
stable~\cite{Chubukov}. Our parameters are far away from these
critical values, and hence the classical ground state, without
dilution, is N\'eel ordered.

\paragraph{Diluted case\\}


One might have expected that the combination of frustration,
brought about by $J_2$ and $J_3$ and site dilution would trigger
an instability in favor of a local Villain canting of the
spins~\cite{Villain-ZPhysB}, leading ultimately to a two
dimensional Heisenberg spin glass before $x_p$ is reached~\cite{Gawiec1}.
\begin{figure}[ht]
\begin{center}
\scalebox{0.7}[0.7]{\includegraphics{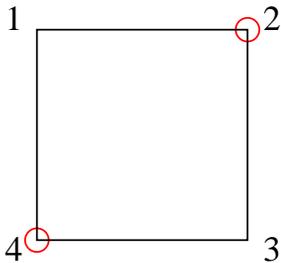}} \caption{Particular
dilution configuration.
In this example, sites $2$ and $4$ are removed by dilution.
} \label{trouss}
\end{center}
\end{figure}

\noindent However, as alluded to in
the discussion below Eq.~(\ref{Jij-dil}),
such locally Villain canted states do not occur
in the model considered here, where all effective magnetic
interactions derive from electronic processes involving nearest
neighbor hopping. Hence, as we saw in the previous section, for
the configuration of diluted sites shown in Fig. \ref{trouss} 
the second neighbor
interaction, $J_{13}$, 
between sites 1 and 3 is destroyed by the dilution 
of sites 2 and 4.
As a consequence, as long as the critical ratios for the $J_2/J_1$,
$J_3/J_1$ or $J_c/J_1$ 
for destroying two sublattice collinear N\'eel order
are not reached, there are no spins
coupled by dominantly {\it random} frustrating interactions, $J_2$,
$J_3$ or $J_c$, as can be verified from studying  Eqs.
(\ref{J1},\ref{J2},\ref{Jc}).

We therefore conclude that the classical ground state
of $H_{\rm s}^{(4)}$ in Eq.~(\ref{Hs4}) for $t/U=1/8$ on the
percolation cluster is a N\'eel configuration for all
concentrations below the percolation threshold. From this, one can
immediately see the importance of the site percolation threshold
in this problem: within the model considered, that is the
site-diluted one-band Hubbard model of Eq.~\ref{HH}, the only
accessible classical ground state is N\'eel ordered all the way to
the percolation threshold $x_p$. Hence, any reduction in the range
of stability of the N\'eel ground state is due uniquely to quantum
fluctuations and is not due to (classical) random frustration
effects. This conclusion is explicitly verified 
post factum
within the real space
spin wave calculation: any instability towards a non-collinear
ground state would be detected as a negative eigenvalue of the
Hessian matrix leading to complex eigenfrequencies. No such
instabilities were detected in more than the ten thousand
realizations of disorder considered in this work.

We note, however, that La$_2$CuO$_4$ is only approximately described
by the one-band Hubbard model with nearest neighbor hopping only.
For instance we have recently shown that one can achieve a
quantitative improvement to the fitting of the spin wave
excitation spectrum measured by Coldea {\it et al.}~\cite{coldea01} by
including direct further neighbor hopping constants $t'$ and
$t''$ ~\cite{Delannoy2}. Such
direct hops could change the above results, leading to canted
classical ground 
states~\cite{Villain-ZPhysB,Kinzel,Saslow1,Saslow2,Vannimenus,Gawiec1}
before the percolation threshold is reached ($x_c<x_p$).
%

\subsection{Elementary excitations of a diluted spin $1/2$ system}

\subsubsection{Method}

The introduction of site dilution destroys translational invariance, which excludes the
use of Fourier space for calculating the spin-wave excitations.
Hence, we closely follow the method introduced by Walker and
Waldstedt \cite{ww} to study excitations in Heisenberg spin
glasses. Other recent studies of site-diluted $S=1/2$ Heisenberg
antiferromagnets have followed a similar
approach~\cite{Castro,Beach}. We first summarize this method for
the simplest case of nearest neighbor exchange, $J_1$ only, with
Hamiltonian
\begin{equation}\label{nnhdilue}
H=\frac{1}{2}\sum_{\langle i,j\rangle}J_1(i,j) \bm{S}_i\cdot \bm{S}_j.
\end{equation}
As we know the  classical ground state of the system, we can
define for each site $i$, a unit vector ${\bm n}_i^0$ pointing in
the direction of the classical spin ${\bm S}_i$ in this state.
Note that in Eqs.(\ref{Mag},\ref{one},\ref{one-tilde})
a unique global quantization axis in the lab frame, $\hat z$ , was used
to define ${\widetilde M}_{\rm s}$ and $M_{\rm s}$.
Henceforth, we label the spin components in terms of the
projection of ${\bm S}_i$ along the axis of a local right handed 
frame. We do so to keep with the original notation of Ref.~[\onlinecite{ww}], 
from which we borrowed the method we use here.
Let $\{{\bm x}_i,{\bm y}_i, {\bm n}_i^0\}$  be an orthogonal triad of
unit vectors and let  ${\bm p}_i^+$ and ${\bm p}_i^-$ be vectors
defined by
\eqn{\begin{array}{ccl}
{{\bm p}_{i}^{+}} &= & \f{{{\bm x}_i}+i{{\bm y}_i}}{\sqrt{2}},\\
\\
{{\bm p}_i^-}     &= & \f{{{\bm x}_i}-i{{\bm y}_i}}{\sqrt{2}}.
\end{array}}
We also introduce spin deviation (boson creation and annihilation
operators), $a_i$ and $a_i^\dagger$, defined by
\begin{eqnarray}\label{def}
{\bm S}_i \cdot {\bm n}_i^0 & = & S-a_i^{\dagger}   \,    a_i   \;\;\; ,   \nonumber   
\\
\\
{\bm S}_i \cdot {\bm p}_i^+
& =& \sqrt{2S}    {\left
 [1-\frac{a_i^{\dagger}\, a_i}{2S}\right ]}^{\frac{1}{2}} a_i
\;\;\; , \\
%
{\bm S}_i \cdot {\bm p}_i^- & =& \sqrt{2S}a_i^{\dagger}{\left
[1-\frac{a_i^{\dagger} \,a_i}{2S} \right ]}^{\frac{1}{2}},
%
\nonumber
\end{eqnarray}
where the spin components are defined with respect to the local
basis set, $\{{\bm x}_i,{\bm y}_i, {\bm n}_i^0\}$. With
Eq.~\ref{def} and the definition of ${\bm p}_i^{\pm}$, we can
rewrite the Hamiltonian Eq.~\ref{nnhdilue} to order $O(S)$ as:
%

\begin{equation}
\begin{split}
H & =    \frac{1}{2} S^2 \sum_{\langle i,j\rangle}J_{ij}
{{\bm n}_i^0}\cdot {{\bm n}_j^0}  \\
 & + \frac{1}{2}  S^{3/2} \sum_{\langle i,j\rangle }J_{ij}
\Big[  {{\bm n}_i^0}\cdot {{\bm p}_{j}^{+}}
a_j^{\dag}       +{{\bm n}_i^0}\cdot      {{\bm p}_{j}^{-}}a_j	\\
& \quad +\; {{\bm n}_j^0} \cdot {{\bm p}_{i}^{+}}
        a_i^{\dag}          +
{{\bm n}_j^0}\cdot {{\bm p}_{i}^{-}}  a_i\Big]	\\ 
& +
 \frac{1}{2} S \sum_{\langle i,j \rangle}J_{ij}
\Big[
({{\bm p}_{i}^{+}}
a_i^{\dag} + {{\bm p}_{i}^{-}} a_i) \cdot
({{\bm p}_{j}^{+}}
a_j^{\dag} + {{\bm p}_{j}^{-}} a_j) \\
& \quad -\
{{\bm n}_i^0} \cdot {{\bm n}_j^0} ( a_i^{\dag} a_i + a_j^{\dag} a_j)\Big]
\; .
\end{split}
\end{equation}

By making reference to  the classical ground state, we introduce
$\lambda_i$ defined by
\eqn{\sum_j J_{ij} \bm{n}_j^0	= \lambda_i
\bm{n}_i^0	\label{lambda}\;\; \; .
}
Physically, $\lambda_i$ corresponds to the local staggered
mean-field at site $i$ originating from all the spins ${\bm S}_j$
to which ${\bm S}_i$ is coupled. This change of variables makes
the second term of the Hamiltonian vanish. 
We keep only the leading quantum correction to the classical term 
$\frac{1}{2}S^2 \sum_{\langle i,j \rangle} {{\bm n}_i^0}\cdot {{\bm n}_j^0}$,
$H_2$, quadratic in the $\{ a_i^\dagger,a^{\phantom{\dagger}}_i \}$:
%
\eqn{
\begin{array}{ccl}
H_2 & =  & \displaystyle S \left\lbrack\sum_i \lambda_i a_i^{\dag} \,
a_i\right. \\&&\displaystyle \left.\- \frac{1}{2} \sum_{\langle i,j \rangle}
J_{ij}({{\bm p}_{i}^{+}} a_i^{\dag} + {{\bm p}_{i}^{-}}
a_i)\cdot({{\bm p}_{j}^{+}} a_j^{\dag} + {{\bm p}_{j}^{-}} a_j)\right\rbrack
\end{array}
}
The quantum-mechanical equations of motion are:
\begin{eqnarray}
-i\frac{da_i^{\dag}}{dt}=\left[H_2,a_i^{\dag}\right]  ,    \nonumber\\     \\
-i\frac{da_i}{dt}=\left[H_2,a_i\right], \nonumber
\end{eqnarray}
which can be written:
\begin{eqnarray}
\f{da_i}{dt}=-i\left(\sum_j    Q_{ij}a_j^{\dag}    \,    +   \,    \sum_j
P_{ij}a_j\right),   \nonumber  \\   \\  \f{da_i^{\dag}}{dt}=i\left(\sum_j
Q_{ij}^{*}a_j \, + \, \sum_j P_{ij}^{*}a_j^{\dag}\right), \nonumber
\end{eqnarray}
where $P_{ij}={\lambda_i}\delta_{ij} - {J_{ij}}({{\bm p}_i}^{+}
\cdot {{\bm p}_j}^{-}) $
     and $Q_{ij}=-
{J_{ij}}({{\bm p}_i}^{+} \cdot {{\bm p}_j}^{+}) $. We use a vector
representation  for the  operators $a_i$ and $a_i^{\dag}$; that is
$a$ and $a^{\dag}$ are $N$-dimensional vectors whose components
are $a_i$ and $a_i^{\dag}$, respectively. As $N$,
the total number of (occupied) magnetic sites,
 is configuration
dependent, so are all vectors and matrices in the following
discussion. We write
\begin{equation} \label{cPQ}
\f{d}{dt}\left(\begin{array}{c} a \\ a^{\dag}
\end{array}\right)\; =\; i \, \,
\left(\begin{array}{cc}
-P & -Q \\ Q^{*} & P^{*}
\end{array}\right)
\;
\left(\begin{array}{c} a \\ a^{\dag}
\end{array}\right),
\end{equation}
where we refer to 
$P$ and $Q$ as the ``interaction matrices'' of order $N\times N$. 
We can also write
\begin{equation}
H_2=S\left(a^{\dag} \;a\right)\; {\widetilde{H}}\;
\left(\begin{array}{c} a
\\ a^{\dag}
\end{array}\right)
\end{equation}
where the Hamiltonian matrix $\widetilde{H}$ is defined in the
$2N\times 2N$ phase space to be:
\begin{equation}\label{int-matrix}
\widetilde{H}=\left(\begin{array}{cc} -P & -Q \\ Q^{*} & P^{*}
\end{array}\right).
\end{equation}
In order to diagonalize $\widetilde{H}$, we perform a Bogoliubov
transformation that introduces new boson operators $d$ and
$d^{\dag}$, as follows:
\begin{eqnarray}
a        & = &  g^{*}\,d+f\,d^{\dag}  \nonumber \\ \\
a^{\dag}  & = & f^{*}\,d+g\,d^{\dag}  \nonumber 
\end{eqnarray}
$f$ and $g$ are $N \times  N$ matrices that must satisfy the
boson commutation rules:
\begin{eqnarray*}
g^*\,f^{T} - f\,g^{\dag} & = & 0 \\
g^*\,g^{T} - f\,f^{\dag} & = & 1    \;\;\;  ,
\end{eqnarray*}
%
%
where $f^{T}$ is the transpose matrix of $f$. We can also write
these relationships in matrix representation:
\eqn{
\widetilde{E}\;\widetilde{I}\;\widetilde{E}^{\dag}\;
=\;\widetilde{I}
\label{boson}
}
where
\begin{equation}\label{gfi}
\widetilde{E}=\left( \begin{array}{cc} g^* &f\\ f^*&g
\end{array}\right),\;\;\;
\widetilde{I}=\left( \begin{array}{cc} -1 &0\\ 0 & 1
\end{array}\right),
\end{equation}
are of dimension $2N\times 2N$.

The  aim  of the  Bogoliubov  transformation  is  to
diagonalize Eq.(\ref{cPQ}). Consequently we require
\begin{equation}
\f{d}{dt}
\left(\begin{array}{c}
d\\ d^{\dag}
\end{array}\right)
\,\,=\,\,
i \left(\begin{array}{cc}
-\Omega & 0\\ 0 & \Omega
\end{array}\right)
\left(\begin{array}{c}
d\\ d^{\dag}
\end{array}\right)
\end{equation}
where  $\Omega$ is  a   diagonal  matrix  of eigenfrequencies.
Using (\ref{cPQ}) and (\ref{gfi}) one obtains:
\begin{align}
& \f{d}{dt}
\left(\begin{array}{cc}
g^*& f\\ f^* & g
\end{array}\right)
\left(\begin{array}{c}
d\\ d^{\dag}
\end{array}\right)  \nonumber
\\
&\qquad \qquad = i 
\left( \begin{array}{cc}
g^* &f\\ f^*&g
\end{array}\right)
\left(\begin{array}{cc}
-\Omega & 0\\ 0 & \Omega
\end{array}\right)
\left(\begin{array}{c}
d\\ d^{\dag}
\end{array}\right)  \nonumber
\\
& \qquad \qquad =  i 
\left( \begin{array}{cc}
-P &-Q\\ Q^*&P^*
\end{array}\right)
\left( \begin{array}{cc}
g^* &f\\ f^*&g
\end{array}\right)
\left(\begin{array}{c}
d\\ d^{\dag}
\end{array}\right). 
\end{align}
Hence, the equation we ultimately have to solve is:
\begin{equation}
\widetilde{E}D= \widetilde H \widetilde{E},
\end{equation}
where we have defined the complete matrix of eigenvalues:
\begin{equation}\label{impor}
D= \left(
\begin{array}{cc} -\Omega&0\\ 0 & \Omega
\end{array}\right)  \;\;\; .
\end{equation}
%
%
%
With this method, we can calculate the zero point quantum spin
fluctuations to order $1/S$, and hence the expectation value for
the spin on occupied site $i$:
\begin{equation}\label{def2}
\moy{ S_i^z}\= S-{\moy{a_{i}^{\dag}a_i}} = S-\sum_{\nu}|f_{i\nu}|^2.
\end{equation}
With the expectation value $\moy{S_i^z}$ now
defined in terms of $|f_{i\nu}|^2$, one can calculate
the staggered magnetization, defined in either Eq.~(\ref{one})
for the finite $t/U$ Hubbard model or Eq.~(\ref{one-tilde})
for the $t/U\rightarrow 0$ Heisenberg model.
Formally speaking, in a thermodynamically large system,
spins that reside on finite-size clusters
and which are connected to the percolating cluster do not
participate to the symmetry breaking nor do they
contribute to the average bulk staggered magnetization.
Hence, to capture that physics in the present problem,
and to proceed numerically,
numerically, we first identify 
for a given realization of disorder, a percolating cluster of
sites connected via nearest neighbor hopping. For each spin
on the percolating cluster, $\moy{S_i^z}$ is determined  from
Eq.~(\ref{def2}), summed over, and normalized
by $N$, the total number of sites
for that realization of disorder (percolating and not), to give
$M_{\rm s}$ in Eq.~(\ref{one}) (henceforth
denoted $M$). One then repeats the calculation
for many dilution configurations for a given $x$,
performing a disorder average and
obtaining both the averaged staggered magnetization
on the percolating cluster, $[M(x)]_{\rm perc}$, or the
{\it bulk} staggered magnetization, $[M(x)]$,
averaged over all magnetic sites in the sample.
We stress that, while the staggered magnetization
on the percolating cluster, $[M(x)]_{\rm perc}$  
is the most relevant quantity for the numerical study, 
it is the average staggered magnetization
over all Cu magnetic sites in the system, percolating and not, $[M(x)]$,
which is accessible to experiment, and which is displayed in Fig.~\ref{science}.

%

In the presence of interactions beyond $J_1(i,j)$,
the only change in the details of the above method
occur in the matrix elements of  $P$ and $Q$.
The form of these matrices, taking into account the second ($J_2$),
third ($J_3$) and ring ($J_c$) exchange interactions is discussed next.

\subsubsection{Calculation of the $P$ and $Q$ matrices}

\paragraph{$J_1$: first NN\\}

In this case the quadratic Hamiltonian reads:
\eqn{H_2(J_1) = S\sum_{\langle i,j \rangle} J_1(i,j)\left( a_i^{\dag}a_i +
a_j^{\dag}a_j  - a_i a_j -
a_i^{\dag}a_j^{\dag}\right).}
The $P$ and $Q$ interaction matrices  then have the following form:
\eqn{\begin{tabular}{cc}
$P\=\left(\begin{array}{cccc}
\lambda_1 \\
& \ddots \\
& & \ddots \\
& & & \lambda_{L^2}
\end{array}\right)$, & 
$Q\=\left(\begin{array}{cccc}
0\\
& \ddots & -J_1 \\
&-J_1  & \ddots \\
& & & 0
\end{array}\right)$\end{tabular} }
where $\lambda_i$ is defined in Eq. (\ref{lambda}), and
$Q$ is a symmetric matrix $Q_{ij}\=-J_1(i,j)$.\\

\paragraph{$J_2$: Second NN\\}

We have:
\eqn{H_2(J_2) = -S\sum_{\langle \langle i,j \rangle \rangle} J_2(i,j)\left(
a_i^{\dag}a_i + a_j^{\dag}a_j  -  a_i^{\dag}a_j
- a_j^{\dag}a_i \right),}
which leads to the following additions to the $P$ and $Q$
matrices:
\begin{align}
P^{(2)} &= \left(\begin{array}{cccc}
\lambda_1^{(2)} \\
& \ddots & -J_2 \\
& -J_2 & \ddots \\
& & & \lambda_{L^2}^{(2)}
\end{array}\right), \nonumber
\\ 
Q^{(2)} &=\left(\begin{array}{cccc}
0 & & &  0\\
\vdots& \ddots & & \vdots \\
\vdots&  & \ddots & \vdots \\
0& & & 0
\end{array}\right)
\end{align}
$\lambda_i^{(2)}$ is defined in a similar way to $\lambda_i$:
 \eqn{\sum_{\langle \langle j \rangle \rangle} J_2(i,j) 
\bm{n}_j^0 =
\lambda_i^{(2)}
\bm{n}_i^0
,\label{lambda2}}
where $\langle \langle j \rangle \rangle$ indicates a sum over the second neighbors of site $i$.

\paragraph{$J_3$: Third NN\\}

\eqn{H_2(J_3) = -S\!\!\sum_{\langle \langle \langle i,j \rangle \rangle \rangle} \!\!J_3(i,j)\left( 
a_i^{\dag}a_i + a_j^{\dag}a_j  -  a_i^{\dag}a_j -
a_j^{\dag}a_i \right),}
hence the expression for the $P$ and $Q$ matrices are modified by:
\begin{align}
P^{(3)} &=\left(\begin{array}{cccc}
\lambda_1^{(3)} \\
& \ddots & -J_3\\
& -J_3 & \ddots \\
& & & \lambda_{L^2}^{(3)}
\end{array}\right), \nonumber
\\
Q^{(3)} &=\left(\begin{array}{cccc}
0 & & &  0\\
\vdots& \ddots & & \vdots \\
\vdots&  & \ddots & \vdots \\
0& & & 0
\end{array}\right)
\end{align}
with $\lambda_i^{(3)}$ defined by:
\eqn{\sum_{\langle \langle \langle j \rangle \rangle \rangle} J_3(i,j)
\bm{n}_j^0 = \lambda_i^{(3)}
\bm{n}_i^0
,\label{lambda3}}
where $\langle \langle \langle j \rangle \rangle \rangle $ indicates the third neighbors of site $i$.

\vspace*{0.2cm}

\paragraph{$J_c$: Ring exchange interaction\\}

To first order in $1/S$, the four spin terms appearing in the
Hamiltonian are decoupled into 
bilinear products of $a_i^\dag a_j$.
That is, to order $1/S$, the net effect of
the ring exchange is to simply renormalize
the $J_1$ and $J_2$ interactions~\cite{coldea01,Toader,Delannoy}.
 The contribution of the ring
exchange terms to the quadratic Hamiltonian is thus:
\begin{widetext}
\begin{equation}\begin{split}
H_2(J_c) &= \displaystyle -S^3
\sum_{\langle i,j,k,l \rangle } J_c(i,j,k,l) \left[ \left( a_i^{\dag}a_i
+ a_j^{\dag}a_j + a_k^{\dag}a_k + a_l^{\dag}a_l\right)
+ \left( a_i^{\dag}a_k + a_k^{\dag}a_i + a_j^{\dag}a_l 
+ a_l^{\dag}a_j \right)\right. \nonumber 
\\
& \left. \qquad\qquad -\ \left( a_i^{\dag}a_j + a_j^{\dag}a_i +   a_i^{\dag}a_l + a_l^{\dag}a_i 
+  a_j^{\dag}a_k + a_k^{\dag}a_j + a_k^{\dag}a_l + a_l^{\dag}a_k \right) \right],
\label{reduc}
\end{split}\end{equation}
\end{widetext}
where $J_c(i,j,k,l)=\epsilon_i \epsilon_j \epsilon_k
\epsilon_l J_c$. 
The elements of the  Bogoliubov transformation matrices $g$
and $f$ are thus modified by the configuration dependent renormalization
of the first and second neighbor
exchanges. In zero dilution, $J_1$ and
$J_2$ are renormalized to~\cite{coldea01,Toader,Delannoy}:
 \eqn{\left\{\begin{array}{cclcl}
J_1^{\rm eff} & = & J_1 - 2 S^2 J_c & = & J_1 - \f{J_c}{2}  \\
\\
J_2^{\rm eff} & = & J_2 - S^2 J_c  & = & J_2 - \f{J_c}{4};
 \end{array}\right.\label{reduc2}}

\section{Algorithmic Considerations}

In order to obtain the quantum magnetization corrections in the
disordered lattice, we have to solve the eigenvalue problem
described in Eq. (\ref{impor}). Results in the thermodynamic limit
are estimated by doing a finite size scaling analysis for
different system sizes. For each value of size and dilution, we
generate many  realizations of disorder after which we perform
successively the disorder average and the finite size scaling to
the thermodynamic limit. Our algorithm is organized as follows
for each value of the system size and dilution:

\begin{itemize}
 \item Generation of the diluted  lattice
and computational identification of the percolating cluster.
 \item Calculation of the $P$ and $Q$ matrices (Eq.~(\ref{cPQ})).
 \item Diagonalization of the matrix using Lapack routines.
\end{itemize}

For a system of linear size $L$ and dilution concentration $x$, for each
site, we generate a random number $r$ between $0$ and $1$. The
site is considered as removed if  $r \geqslant (1-x)$. For each
realization of disorder, for which the number of sites
$N(L^2,x)$ is different, we first construct the percolating
cluster. To do this, the undiluted sites are labeled from $1$ to
$N$. Starting from site $1$, with coordinates $(\alpha, \beta)$,
we verify if the neighbors $(\alpha\pm 1, \beta)$, $(\alpha, \beta
\pm 1)$ are occupied. If yes the label of the site is changed to
$1$. Moving to one of these sites the procedure is repeated. If the
cluster $1$ terminates, the next cluster takes the number of the
first occupied site encountered. Once all sites have been visited,
the procedure is repeated taking an arbitrary starting point. If
neighboring sites are occupied the indices of the two sites take
the lowest of the two values. The procedure is repeated until no
further evolution occurs. For the biggest cluster we then check
for the existence of percolating pathways along the $x$ and $y$
direction. If a percolating cluster exists, then 
the matrix $\widetilde H$ (\ref{int-matrix}) is constructed.

The diagonalization of $\widetilde H$
is performed using a fortran 77 Lapack double precision set of routines:

\begin{itemize}
\item {\bf DGEHD2} computes Hessenberg reduction of the  $\widetilde H$ matrix.
\item {\bf DORGHR} and {\bf DHSEQR} lead to the Shur factorization.
\item {\bf DTREVC} gives the eigenvectors of the  $\widetilde H$  matrix.
\end{itemize}


From the results of the Lapack routines 
we first construct a matrix of eigenvectors $E$ of $\widetilde H$.
%
We order the columns of $E$  
so that the first column is an eigenvector corresponding to the lowest
eigenfrequency, and the last column is an eigenvector
corresponding to the highest eigenfrequency of $H$.

The matrix $E$
is thus defined up to  the subspaces of degenerate
eigenvectors and the matrix $D$ in Eq.~(\ref{impor}) is the diagonal
matrix of its eigenfrequencies:
%
\eqn{ E\;D\; E^{-1}\;=\; \widetilde{H} \;\;\; . }
However, knowledge of $D$ and $E$ does not completely determine
the problem. In order to establish the elements of the Bogoliubov
transformation we must construct from $E$ the matrix
$\widetilde{E}$ that satisfies both the relation (\ref{boson}),
coming from the boson commutation relations and the eigenvalue
Eq.~(\ref{impor}). That is:
\eqn{\widetilde{E}\;\widetilde{I}\;\widetilde{E}^{\dag}\;
=\;\widetilde{I}\,\,\,\,\mbox{and}\,\,\,\,\widetilde{E}\;D\;\widetilde{E}^{-1}\;=\;
\tilde{H}}
We find $\widetilde{E}$ through the application of a
transformation:
\eqn{\widetilde{E}\;=\;E\;db,}
where $db$ is a block-diagonal matrix.
Using the commutation relation (\ref{boson}) one finds
\eqn{
 db\;\widetilde{I}\;db^{\dag}=\left(M\right)^{-1}}
with
\begin{equation}\label{Mi}
M=E^{\dag}\;\widetilde{I}\;E.
\end{equation}
$M$ is a Hermitian matrix obtained from the Lapack routines.
It is block diagonal, with blocks $M_i$ of size $p_i\times p_i$,
corresponding to a subspace of degenerate eigenvalues of
$\tilde{H}$, of dimension $p_i$. The transformation matrix $db$ is
therefore also block diagonal, with corresponding blocks $A_i$. If
the matrix, $E_i$, represents the subspace of eigenvectors, of
dimension $2N\times p_i$, the transformation gives
$\widetilde{E}_i=E_i A_i$.

To find $A_i$, we need to solve
\begin{equation}\label{pb}
\pm A_i\;A_i^{\dag}\;=\;M^{-1}_i,
\end{equation}
where the sign $\pm$ depends on which sector of the eigenvalue
matrix $D$ in Eq.~(\ref{impor}) the subspace belongs. The
blocks $M_i$ are first inverted and then diagonalized
\begin{equation}
  M_i^{-1} = K_i\,D_i\,K_i^{-1}.\label{DK}
\end{equation}
The matrix $D_i$ contains either positive or negative eigenvalues,
depending on the sign of the eigenvalue of the subspace of
$\tilde{H}$. From this we find the diagonal matrix  $\sqrt{\pm
D_i}$, where the sign $\pm$ is chosen so that the square root is
defined.

$A_i$ is finally found from:
\eqn{A_i\;=\;K_i\,\sqrt{\pm D_i}\,K_i^{-1},\label{DK2}}
In this block diagonal procedure the subspace corresponding to the
Goldstone modes is explicitly excluded. All other operations are then
mathematically well defined~\cite{JY-thesis}
 and the  Bogoliubov transformation is
completely determined. A further summary of the calculation
procedure can be found in Appendix \ref{Calc}.

%
%

\section{Results}

We have calculated the quantum fluctuations of the magnetic ground
states for a diluted spin system for two situations:

\begin{enumerate}
\item  $t/U \rightarrow 0$, Heisenberg limit. 
The result are
compared with those obtained by Mucciolo {\it et
al.}~\cite{Castro} who used a similar linear spin wave calculation.
The motivation here is to validate the two sets of results against
each other and to quantify finite size effects and statistical
errors. In this limit the additional terms in the magnetization
operator (\ref{effmag}) leading to the inequality (\ref{one-tilde})
are zero. 
This calibration allows us to confirm that there is
indeed a discrepancy between the results from
$1/S$ calculations~\cite{Castro} 
and those from quantum Monte Carlo simulations~\cite{Sandvik}
 in the Heisenberg limit.

\item $t/U\= 1/8$. This $t/U$ value is close to the one found
for La$_2$CuO$_4$ by Coldea {\it et al.}~\cite{coldea01} 
($t/U\simeq 1/7.35$)~\cite{Note-ttt}. By using this value, we
can begin an investigation of the effects of further neighbor and ring exchange
interactions in the experimentally relevant situation of 
Cu substitution by Mg and Zn in La$_2$Cu$_{1-x}$(Mg/Zn)$_x$O$_4$.

\end{enumerate}

\subsection{${\mathbf\mathit t/U}\= 0$: diluted Heisenberg model}

The presence of dilution introduces statistical fluctuations in
the ground state magnetization due both to configurational
variations for fixed number of magnetic sites and to the variation
in the number of magnetic sites from one configuration to another. To combat
this, we perform an average over a number of  disorder configurations, 
${\cal N}_0$,  that increases with the dilution.
We chose ${\cal N}_0$ to be the integer closest to $5000$ times
$x$, where $x$ is the concentration of missing magnetic sites.
Averaging over the disorder we
define the average number of sites for a given concentration:
\eqn{\bar{N} \,\equiv\, \f{1}{\cal N}_0 
\sum
N(L^2,x) \= (1-x)L^2,\label{moyennage}}
where $N(L^2,x)$ is the number of sites for a system 
of size $L$ and concentration $x$ for a
specific disorder realization.
 System sizes were studied from $L^2\=10\times 10$ to
$L^2\=34\times 34$. A detailed discussion of the 
various contributions to the statistical
errors can be found in Appendix \ref{errors}.


The ground state magnetization is estimated by extrapolating the
finite size results to the thermodynamic limit. In order to do
this we proceed in two steps:

\begin{itemize}

\item
 Firstly we determine
 the staggered magnetization for the
 sites on the percolating
 cluster for each realization of disorder; from which the
 staggered magnetization averaged over all magnetic sites for a
 specific realization of disorder is obtained.
 We then make a disorder average over many realizations, calculating both the
 disorder averaged staggered magnetization on the percolating
 cluster, $[M(x,L)]_{\rm perc}$, and the experimentally relevant disorder averaged 
 bulk staggered magnetization, $[M(x,L)]$. The errors on these measures are
 estimated as explained in Appendix \ref{errors}.


\item This process is repeated for different system sizes for a given 
$x$ and the results are extrapolated to the thermodynamic limit by making a 
least squares fit of the form~\cite{Huse}:
\begin{equation}
[M(x,L)]_{\rm perc} \= [M(x)]_{\rm perc} \+ \f{a}{L} \+ \f{b}{L^2} ~.
\label{Huse-scaling}
\end{equation}
The same procedure is used for  $[M(x,L)]$.

\end{itemize}

\begin{figure}[ht]
\begin{center}
\includegraphics[width=8cm,height=7cm]{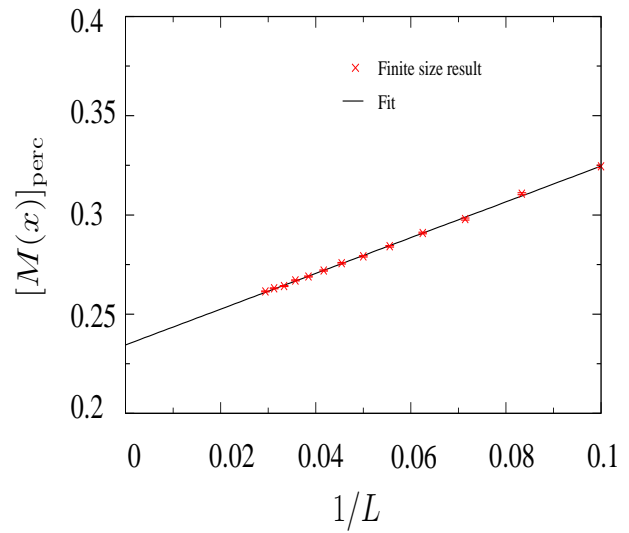}
\caption{Staggered magnetization for the
$t/U \rightarrow$ Heisenberg model 
for $x=20\%$ and $L\in [10,34]$.
}\label{res2_fig}
\end{center}
\end{figure}

\begin{figure}[ht]
\begin{center}
\includegraphics[width=8cm,height=7cm]{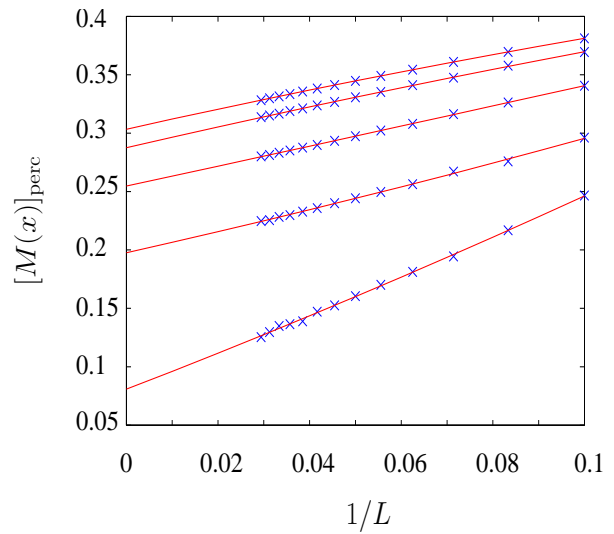}
\caption{Evolution of $[M]_{\rm perc}$ 
for the $t/U \rightarrow$ Heisenberg model 
with $L$ for different values
of dilution $-$ from top to bottom: $x=0\%$, $x=6\%$, $x=16\%$,
$x=26\%$ and $x=36\%$.}\label{res3_fig}
\end{center}
\end{figure}

As an example, we show results for $x=0.2$ in Fig.
\ref{res2_fig} where we plot the magnetization
$[M(x,L)]_{\rm perc}$  against $1/L$. As one
can see, the statistical noise on the data is small and is
consistent with the size of the error bars estimated in 
Appendix \ref{errors}. The magnetization extrapolates linearly to the
thermodynamic limit in $1/L$ to an excellent approximation,
allowing a high precision estimate for $[M]_{\rm perc}$:
 \eqn{[M(x=0.2)]_{\rm perc} \=0.236 \pm 0.001}
We note that between $L=10$ and the biggest
system size studied, $L=34$, $[M]_{\rm perc}$ varies by
over $30\%$. This substantial variation confirms the need for such
a finite size scaling procedure here. Results for different
values of $x$ are shown in Fig.~\ref{res3_fig}. For the system
sizes studied, the size dependence is very nearly linear in $1/L$
for all $x$.
One can also notice that the slope, $a$, is
almost independent of $x$ until the percolation threshold,
$x_p=0.41$ is approached, at which point it increases with finite
size effects becoming progressively more important. This evolution
is not inconsistent with the critical nature of the percolation
threshold and the question as to whether $[M]_{\rm perc}$,
determined via the $1/S$ method,
goes continuously to zero or jumps discontinuously to zero at
$x_p$ is an intriguing one. 
On the other hand, it is found from quantum Monte Carlo simulations that
$[M]_{\rm perc}$ has a
discontinuous jump at $x=x_p$ ~\cite{Sandvik}.
However, this question is not the main
focus of the paper and to do it justice would require a more
extensive and dedicated study near $x_p$. Here we simply remark
that $[M]_{\rm perc}$ extrapolates to small values for
concentrations less than, but near $x_p$.

\begin{figure}[ht]
\begin{center}
\includegraphics[width=8cm,height=7cm]{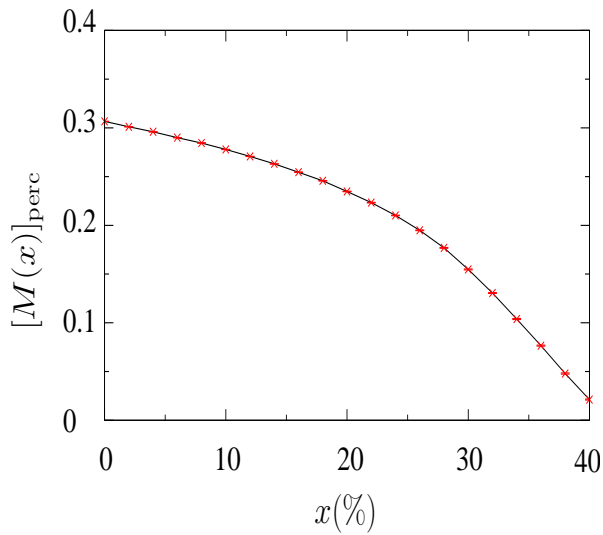}
\caption{Staggered magnetization 
for the $t/U \rightarrow$ Heisenberg model 
on the percolating cluster extrapolated to the
thermodynamic limit.
The solid line is a guide to the eye.}\label{res4_fig}
\end{center}
\end{figure}

Collecting these results, we show the staggered
magnetization for the ground state of the site-diluted Heisenberg
model, as a function of dilution in Fig. \ref{res4_fig}. 
$[M]_{\rm perc}$ goes smoothly 
from the known value for
the undiluted case
in the $1/S$ approximation~\cite{Anderson-sw,Stinchcombe-sw,Manousakis}, 
 $[M]_{\rm perc}\approx 0.31$,
to zero for $x$ very close to the site dilution percolation threshold, $x_p$.

\begin{figure}[ht]
\begin{center}
\includegraphics[width=8cm,height=7cm]{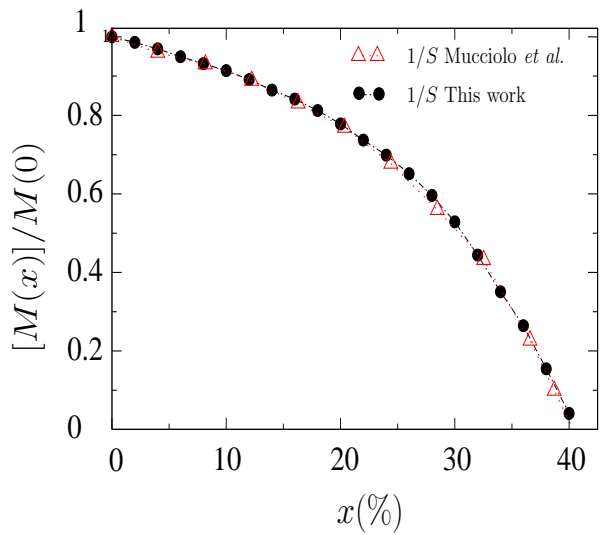}
\caption{Comparison between our data for the ground state (bulk averaged)
staggered
magnetization for the Heisenberg model, normalized by the value 
at zero dilution, and the
data of Mucciollo {\it et al.}\cite{Castro}
The dashed-dotted line is a guide to the eye.}\label{res5_fig}
\end{center}
\end{figure}

In Fig.~\ref{res5_fig}, we compare our result with those obtained
by Mucciolo {\it et al.} \cite{Castro}
for the same model.
The data are normalized by the value
$[M]_{\rm perc}(x=0) \equiv M(0)$.
 There is extremely good
quantitative agreement 
between our results and theirs, providing
strong evidence that the two
methods give correct results
for the $1/S$ method considered.

%
%

\begin{figure}[ht]
\begin{center}
\includegraphics[width=8cm,height=7cm]{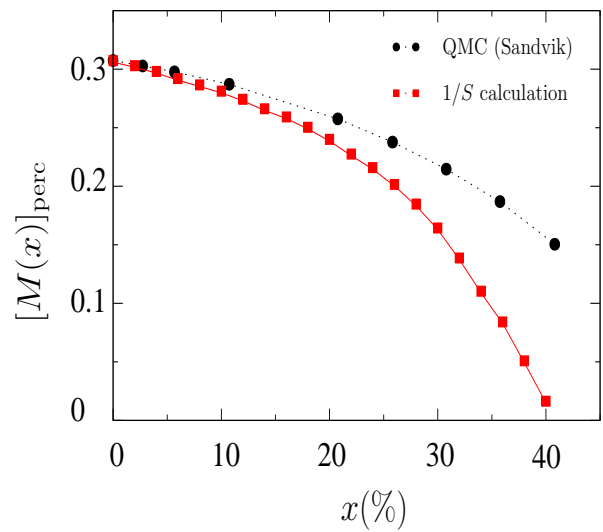}
\caption{Staggered magnetization on the percolating cluster for
the site-diluted Heisenberg  model: comparison between 
QMC (from Ref.~[\onlinecite{Sandvik}]) and $1/S$ spin wave results.
The solid and dashed lines are guide to the eye.}\label{res7_fig}
\end{center}
\end{figure}

It is important here to make a
comparison between our results and those from quantum Monte Carlo
(QMC), which is in principle exact, apart from numerical error.
Such a comparison is made in Fig. ~\ref{res7_fig} where we
show unnormalized data for the magnetic moment on the percolation
cluster from our calculation, compared with the QMC data of
Ref.~\onlinecite{Sandvik}. For zero dilution, the methods give very
similar results. This is expected as it is known that $1/S^2$
contributions to the quantum fluctuations in this case are
identically zero~\cite{Stinchcombe-sw,Igarashi,Igarashi2}, 
meaning that the difference between spin wave
and QMC comes, to leading order, from $1/S^3$ contributions, which
one might expect to be small. Moving away from zero dilution, the
difference between the two sets of results increases
in a monotonic way,
with the
moment from the QMC consistently larger than
that determined from the $1/S$ spin wave calculation.
Hence the comparison explicitly illustrates that the $1/S$ method
over estimates the importance of the quantum fluctuations in the
presence of disorder. In order to understand this quantitative
difference, one 
should investigate
the effects of magnon-magnon interactions and Berry phase
terms, which we do not attempt here.

\begin{figure}[ht]
\begin{center}
\includegraphics[width=8cm,height=7cm]{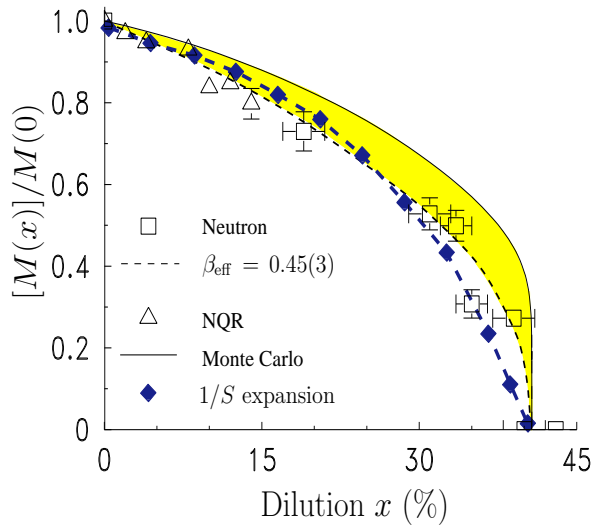}
\caption{Ground state magnetization as a function of dilution for
Mg and Zn doped La$_2$CuO$_4$:
La$_2$Cu$_{1-x}$(Mg/Zn)$_x$O$_4$~\cite{Vajk}, for quantum Monte
Carlo~\cite{Sandvik} for the site-diluted 
square lattice Heisenberg antiferromagnet (SLHAF). 
The dashed line is a guide to the eye parameterized by 
$[M(x)]/M(0)=(1-x/x_p)^{\beta_{\rm eff}}$.
The figure is 
reproduced
from Ref.~[\onlinecite{Vajk}]. Added to this figure is our data
and that of Ref.~[\onlinecite{Castro}] obtained from the numerical $1/S$ 
spin wave analysis of
the site-diluted SLHAF.}\label{res6_fig}.
\end{center}
\end{figure}

The above limitations should be taken into
consideration when comparing data from spin wave calculations with
the experiment. In Fig.~\ref{res6_fig} we add our results to those
previously shown in Fig. 1 for comparison with the experimental
neutron scattering data on
La$_2$Cu$_{1-x}$(Mg/Zn)$_x$O$_4$~\cite{Vajk} and quantum Monte Carlo
simulations~\cite{Sandvik}. The figure allows us to confirm the
conclusion, already made in Ref.~\onlinecite{Castro}, that the extremely
good agreement between experiment and the spin wave calculation is
rather fortuitous: if the Heisenberg Hamiltonian was an adequate
starting point to describe the experimental data, the ``exact''
QMC results would be in better
agreement with the experimental data than the $1/S$ 
spin wave data are.  As can be seen from the
figure, the reverse is true; while the QMC data is consistently
above the experimental curve, the spin wave data lies very close
to it. Hence, although the Heisenberg Hamiltonian is clearly a
good starting point for acquiring an acceptable qualitative
description of Mg and Zn doped La$_2$CuO$_4$ it appears,
on the basis of the results shown in Fig.~\ref{res6_fig},
to be inadequate for a really quantitative description. Further, we
remark that the experimental data are presented such that they are
normalized by
$[M(x=0)]_{\rm perc}$. While the undiluted moment is
$[M(x=0)]_{\rm perc} \approx 0.31$ from QMC and spin wave
calculations,
recent estimates by Lee {\it et al.} \cite{Lee} place the
experimental moment at about $0.25$. Hence, removing the absolute
scale of the magnetic moment improves the impression of a good agreement
of the experiment with the QMC results for the site-diluted 
Heisenberg model for small $x$. 
When plotted on an absolute scale the agreement
between experiment and theory would be less convincing. This is an
important point for the present paper as we continue to work
within the linear spin wave approximation and cannot expect to
account for the contributions beyond $1/S$ linear spin waves which,
following the QMC approach, appear to be important for the dilution problem.
That said, by making improvements to the starting spin-only
description of the Hubbard Hamiltonian, through
higher order terms in the canonical transformation,  we can expect
to improve the comparison with experiment on an 
{\it absolute} $[M(x)]$ scale.

With this in mind we have extended our calculations to order
$t^4/U^3$, which allows us to include second and third neighbor
exchange as well as ring exchange around an elementary plaquette,
and also to include quantum fluctuations from charge
delocalization in the underlying Hubbard model.

\subsection{${\mathbf\mathit t/U}\= 1/8$: on the role of the ring exchange interaction}

When interactions beyond nearest neighbor 
exchange 
are taken into account, two
effects have to be considered. First the transverse spin
fluctuations are modified by the inclusion of the new
interactions 
since these affect the magnon excitation spectrum~\cite{coldea01,Delannoy2}.
Secondly, the charge delocalization induces a
further quantum fluctuation term over and above those from
transverse spin fluctuations. This is the difference between
$\hat M_{\rm s}$ and $\hat {\widetilde M}_{\rm s}$  
in Eq.~(\ref{one})
and which leads to renormalization of the staggered magnetization in a way that
depends on dilution. In this section,  we treat these two
effects separately to quantify their respective importance for
$t/U\=1/8$.


\subsubsection{In the absence of charge mobility renormalization}
\label{No-charge-renorm}

\noindent \underline{Finite size results:}

\begin{figure}[ht]
\begin{center}
\includegraphics[width=8cm,height=7cm]{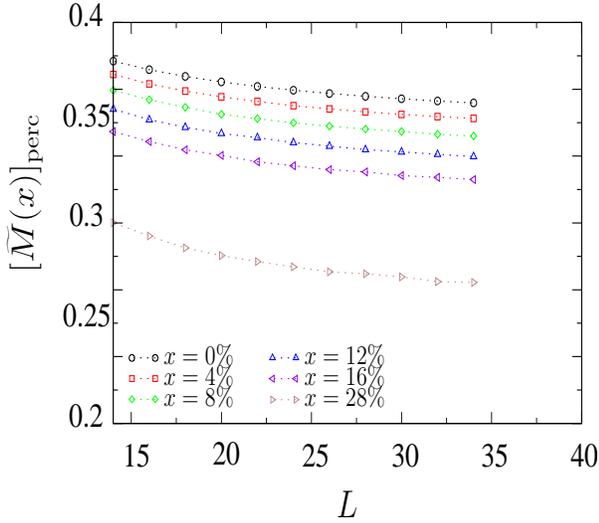}
\caption{Size dependence of the staggered magnetization of the
effective spin-only Hamiltonian $H_{\rm s}^{(4)}$.
 Further neighbor interactions 
are generated through the existence of nearest neighbor 
electronic hopping pathways.
Charge mobility renormalization effects are not included here.
}\label{res11_fig}
\end{center}
\end{figure}

\begin{figure}[ht]
\begin{center}
\includegraphics[width=8cm,height=7cm]{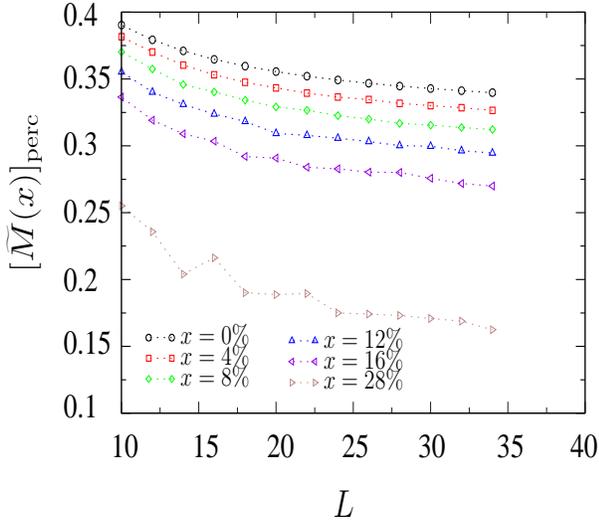}
\caption{Size dependence of the staggered magnetization of a 
``p-model'' Hamiltonian where further neighbor interactions
are not explicitly dependent on the neighbor exchange
hopping pathways.}\label{res9_fig}
\end{center}
\end{figure}

The first point we wish to illustrate here is the importance of 
the modification of the
exchange pathways in the diluted system. 
We have argued in Section
(\ref{pathways}) that dilution does not introduce {\it random} 
frustration at the classical level,
even in the presence of further neighbor spin interactions, 
if these interactions are derived from the
Hubbard model with nearest neighbor hops only. 
In this case, such a longer range
interaction depends on the presence of a nearest neighbor exchange pathway. 
Hence, we do not expect long range
interactions to have a destabilizing effect on 
classical N\'eel order on the percolating cluster.
This can be seen indirectly by comparing the finite size scaling
of our effective spin-only model
with that of a more phenomenological
model. In the latter, which we refer 
to as the ``p-model'', the  further neighbor interactions have
full strength, independently of the existence of a nearest neighbor  exchange path 
created by the electronic hopping processes,
so that they exist even if the
pathway is severed by a non-magnetic defect (i.e. diluted site). 
In the p-model, the bilinear exchange interactions 
$J_2$ and $J_3$ are taken to be
$J_2=4t^4/U^3 \epsilon_i\epsilon_j$
and $J_3=4t^2/U^3 \epsilon_i\epsilon_k$
while $J_c$ is kept to have the same site occupancy dependence
as in Eq.~(\ref{Jc}).
In Fig.~\ref{res11_fig}, we show results for the size dependence of
the staggered magnetization $[{\widetilde M}]_{\rm perc}$
as a function of concentration $x$ of diluted sites.
for the effective spin-only $H_{\rm s}^{(4)}$ in Eq.~(\ref{Hs4})
with $\{J_1,J_2,J_3,J_c\}$ coupling constants as
given in Eqs.~(\ref{J1}) to (\ref{Jc}).
The magnetization is a monotonic function
of  $L$ for all values of $x$.
%
%
This should be compared with Fig.~\ref{res9_fig} where we show
similar data for the p-model.
For large dilution, the statistics are much worse and the
magnetization considerably lower than in the first case. This
indicates the build up of random frustrated plaquettes that
eventually destroy the N\'eel order {\it before} $x_p$ is reached, even
at the classical 
level~\cite{Villain-ZPhysB,Kinzel,Saslow1,Saslow2,Vannimenus,Gawiec1}.

\noindent \underline{Thermodynamic limit}

\begin{figure}[ht]
\begin{center}
\includegraphics[width=8cm,height=7cm]{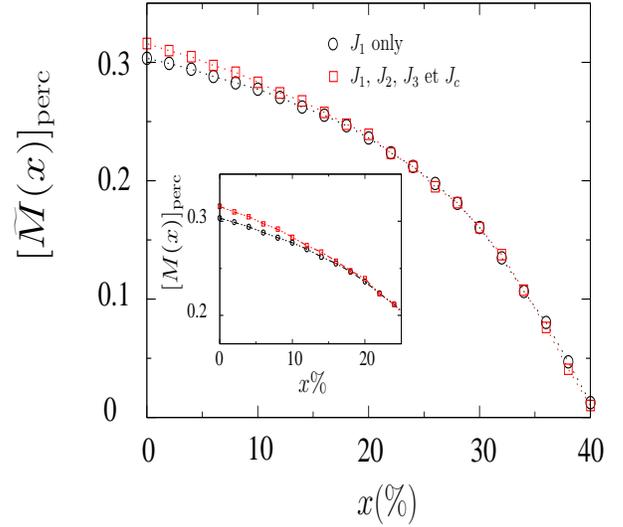}
\caption{Dilution dependence of the staggered magnetization of the
effective
spin-only  Hamiltonian obtained from the Hubbard model - No charge
renormalization effect.
The inset shows a blow-up for $x<20\%$.
}
\label{res12_fig}
\end{center}
\end{figure}

For the effective spin-only Hamiltonian, results are extrapolated to the
thermodynamic limit, using the procedure described in the previous
section and Eq.~(\ref{Huse-scaling}). In Fig. \ref{res12_fig} we show the
ground state magnetization, $[{\widetilde M}]_{\rm perc}$, compared
with the previously shown results from Fig.~\ref{res4_fig} for the 
($t/U\rightarrow$, $J_1$ only) Heisenberg model.

The first thing to notice is that there is very little
difference with the Heisenberg model! 
The second is that the small difference that is
present is towards a {\it higher} ground state magnetization, with the
maximum change occurring at $x\=0\%$. This is because the ring
exchange terms 
decrease the transverse spin 
fluctuations~\cite{Delannoy,Delannoy2}.
As discussed in Refs.~[\onlinecite{Delannoy,Delannoy2}],
this  increase in the
magnetic moment occurs because the ring exchange terms 
in $H_s^{(4)}$  decouple in the $1/S$ expansion into 
effective ferromagnetic second neighbor
two-body exchange terms which 
further stabilizes the two sublattice N\'eel order  by
reducing the transverse spin fluctuations~\cite{Delannoy,Delannoy2} 
(see Eq.~(\ref{reduc2})).
For $x$ greater than about $12\%$ dilution, 
this stabilization effect is largely destroyed 
and the two curves merge up to the percolation threshold. 
This is explained by the fact that, as these
interactions involve more than two sites, they are more sensitive
to dilution than the nearest neighbor terms, and their effect
becomes negligible long before the percolation threshold is reached. 
The effects at high dilution would be very different for 
the p-model (see Fig. \ref{res9_fig}) with frustrating
further neighbor interactions that are independent of the presence
of nearest neighbor exchange pathways. However, in the
context of comparison with experimental results 
on La$_2$Cu$_{1-x}$(Mg/Zn)$_x$O$_4$,
such terms should only appear through electronic hopping over
further neighbors in the Hubbard model. We have recently considered
this problem in the absence of dilution~\cite{Delannoy2},
 but extending this work to include dilution
is beyond the scope of the present study.

\subsubsection{Finite charge mobility renormalization}

\begin{figure}[ht]
\begin{center}
\includegraphics[width=8cm,height=7cm]{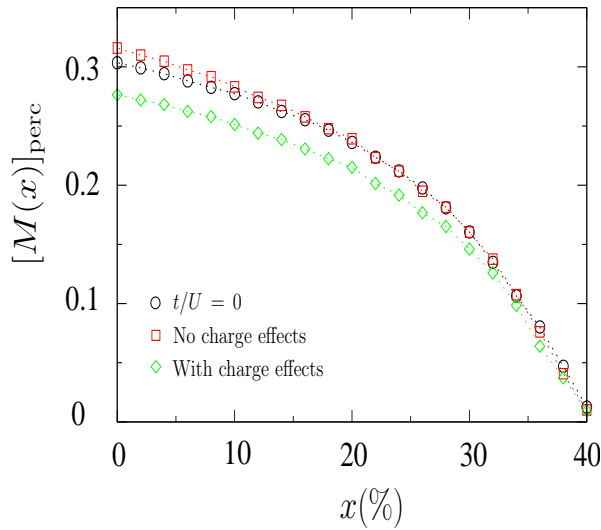}
\caption{Evolution of the staggered magnetization with
dilution for the effective spin-only Hamiltonian with charge
mobility effects included.
The open circles are the data for the Heisenberg ($t/U\rightarrow 0$) model.
The diamonds and squares 
are respectively  the data for the spin-only representation of
the $t/U=1/8$ site-diluted Hubbard model with and without 
finite charge mobility renormalization included. 
}\label{res14_fig}
\end{center}
\end{figure}

The charge mobility, or electron delocalization effect, 
leads to a decrease in the magnetization~\cite{Delannoy,Delannoy2} 
(see Eq.~(\ref{one})). 
However, the
delocalization is also conditioned by the allowed nearest
neighbor electronic hopping pathways  and is consequently also dilution
dependent. The finite size scaling of the magnetization,
as described by Eq.~(\ref{Huse-scaling}), is not
changed qualitatively by this renormalization (not shown here) and
the results are extrapolated to the thermodynamic limit, using 
Eq.~(\ref{Huse-scaling}).
 As shown in Fig.~\ref{res14_fig}, there is a significant decrease in the
magnetization compared with $[ {\widetilde M}]_{\rm perc}$ 
without charge mobility renormalization,
or with the Heisenberg model (see Fig.~\ref{res12_fig}). This difference is
again reduced as the dilution increases.
For $x=0\%$ the decrease is of the order of $14\%$,
whereas it goes down to $9.5\%$ for $x=30\%$ and goes towards zero
at the percolation threshold. We conclude therefore that the
charge delocalization term is a major contribution to the
corrections found by extending, to order $t^4/U^3$, the canonical
transformation of the Hubbard model into an effective spin
Hamiltonian. It is explicitly a property of the Hubbard model and
is not present in a phenomenological spin-only model. It is 
therefore clear that care must be taken when using such phenomenological
spin-only models without directly considering the mobility of the underlying
system of electrons when aiming at obtaining a 
quantitative description and comparison between experiment and a microscopic
theory.
This is the main result of this paper.

\subsection{Experimental considerations}


Figure \ref{res14_fig} illustrates our main
result concerning the comparison with experiment: inclusion of the
charge mobility renormalization factor shifts the scale of
magnetization downwards over the whole dilution range.
Comparing results for zero dilution; for the Heisenberg model, the
ground state moment is $[M_{\rm s}] \approx 0.31$, while
experiment yields $[M_{\rm s}]_{\rm exp}
\approx 0.25$ ~\cite{Lee}. 
Hence a comparison of $[M(x)]$ data not normalized by $M(x=0)$ 
will show the Heisenberg model, either from spin
wave, or from QMC to be {\it above} those from the dilution experiments.
Including hopping processes to order $t^4/U^3$ for $t/U = 1/8$,
a fair estimate for La$_2$CuO$_4$, one
finds~\cite{coldea01,Delannoy2}  $[M_{\rm s}] \approx 0.27$.
This is still above the experimental value, but it is clear that,
taken altogether,
the extra corrections arising from both
transverse spin fluctuations and finite electron mobility away
from the $t/U\rightarrow 0$ Heisenberg limit,
have scaled the magnetization in the right
direction. This is an important result indicating that the
perturbative methods proposed here can describe many of the
magnetic features of La$_2$Cu$_{1-x}$(Mg/Zn)$_x$O$_4$.

\begin{figure}[ht]
\begin{center}
\includegraphics[width=8cm,height=7cm]{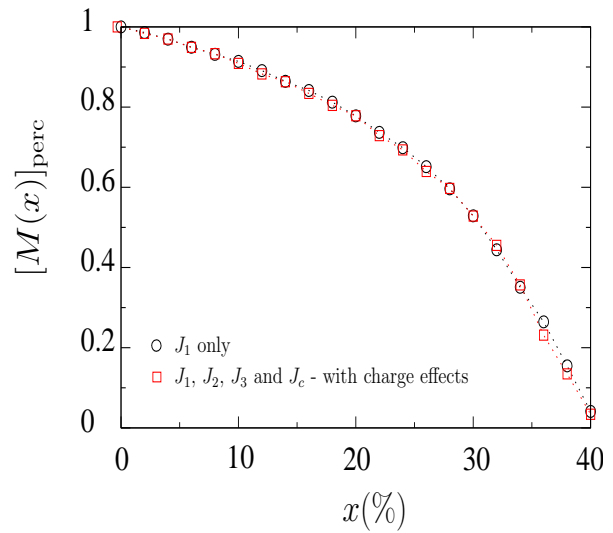}
\caption{Data from Fig. \ref{res14_fig} normalized by the ground
state order parameter at zero dilution, $M(x=0)$.
}\label{res15_fig}
\end{center}
\end{figure}

In Fig.~\ref{res15_fig}, we re-plot the data
of Fig.~ \ref{res14_fig} normalized by the ground state order
parameter at zero dilution, $M(x=0)$. 
The two data sets for the
Heisenberg model and that for the spin model $H_s^{(4)}$
with further neighbor interactions with the
effects of electron delocalization via virtual hopping included,
lie on top of each other. Hence, the improvements brought in by
developing the effective spin description of the Hubbard model up
to four virtual hopping terms, including ring exchange are not
evident when the data are normalized in this way. The data sets
would therefore 
continue to give the same favorable, but fortuitous
comparison with the experimental data of Vajk {\it et al.}~\cite{Vajk},
as seen in Fig.~\ref{res6_fig} and discussed in Ref.~[\onlinecite{Castro}].

\section{Conclusions and perspectives}

\subsection{Conclusions}

In this paper we have investigated the
problem of site dilution in systems described by the half-filled one-band
Hubbard model.  We have extended the canonical transformation technique to
calculate an effective spin Hamiltonian up to order $t^4/U^3$, for
magnetic site dilution $x$. We use a real space spin wave
technique, linear in ($1/S$)  to calculate the dilution dependence of quantum fluctuations
on the staggered magnetization. 
Specifically, we considered two problems.
We first studied the Heisenberg $t/U \rightarrow 0$ limit, comparing our results
with those from quantum Monte Carlo (QMC) studies~\cite{Sandvik}
 on the same model. We confirm,
to a high degree of accuracy, previous
results from Ref.~\onlinecite{Castro}, using a similar  $1/S$  technique.
Hence, our results also confirm 
a systematic deviation between the QMC ~\cite{Sandvik} and
the spin wave results for finite dilution.  
This difference, which is
small for zero dilution, illustrates the dilution dependent generation
of magnon-magnon interactions and Berry phase terms,
both of which are 
 neglected in the $1/S$ spin wave calculation.
By comparing the QMC results and the $1/S$ results, one
concludes that these two effects
 work to stabilize the
semi-classical two sublattice N\'eel order rather than to drive the
system into an exotic quantum phase.  Hence, while the spin wave
technique predicts the magnetic moment  on the percolating cluster
going to zero at, or very close
to the percolation threshold, QMC simulations find a renormalized classical
result, with the moment on the percolating cluster
taking a discontinuous jump to zero at the
percolation threshold.

The second and main objective of our work was to investigate how 
corrections to a site-diluted spin-only Hamiltonian, originating from
a site-diluted one-band Hubbard model affects the dilution
dependence of the N\'eel order parameter (staggered magnetization $[M(x)]$).
Underlying this question, was the goal of obtaining some information
on the capacity of the one-band Hubbard model to
describe data from experiments on 
site-diluted La$_2$CuO$_4$, La$_2$Cu$_{1-x}$(Mg/Zn)$_x$O$_4$ ~\cite{Vajk}.

Within the one-band Hubbard model, the best estimates from fitting to
magnon excitation spectra~\cite{coldea01,Delannoy2} give $t/U \approx
1/8$, placing the system away from the Heisenberg limit and into the
intermediate coupling regime, where higher order electron correlations
need to be taken into account. Integrating out the kinetic degrees of freedom via
a canonical transformation, implemented  to order $t^4/U^3$,
introduces second and third neighbor interactions into the
resulting spin-only Hamiltonian, as well as ring exchange term from
electronic pathways around a closed square plaquette.  
Calculation of the magnetization operator   for the original Hubbard model, 
introduces at this order a charge
mobility term that renormalizes the magnetization
below that 
obtained by considering the (trivial) definition of
the staggered magnetization of a
spin-only Hamiltonian given by Eq.~(\ref{one-tilde}).
 Including all these effects,
and staying within the linear ($1/S$)
 spin wave approximation, we find a reduced
estimate for the moment at zero dilution, $\approx
0.27 \mu_{\rm B}$ compared with $0.31 \mu_{\rm B}$ for the Heisenberg model
~\cite{Delannoy}. As further
neighbor and ring exchange interactions 
mediated by four electronic hops require  unbroken exchange
pathways over length scales greater than nearest neighbor, their
effects disappear well before the percolation threshold is reached.
The net result is therefore that the evolution of the
ground state moment as a function of dilution is qualitatively
similar to that for the Heisenberg model,
disappearing at the percolation threshold in the same way, but with the absolute scale
renormalized {\it downwards} by $10\%$ to $15\%$.
While the $1/S$ method is subject to the limitations described above,
our results  clearly illustrate that for an ultimate detailed and 
 {\it quantitative} understanding
of the role of site dilution in a correlated electron system, such
as La$_2$Cu$_{1-x}$(Mg/Zn)$_x$O$_4$, 
the  charge mobility effects must be taken into account.
Such a description is beyond a spin-only model, decoupled from an
electronic model describing the behavior of the strongly correlated electrons.

\subsection{Perspectives}

To expand on the work presented in this paper,
it would be interesting to carry out further theoretical and numerical studies
using a common calculation scheme for
both the site-diluted Hubbard model expressed in the framework of
a spin-only Hamiltonian with ring exchange 
and the site-diluted
Heisenberg ($t/U\rightarrow 0$) model.
However, in the absence of a solution to the sign problem for frustrated
quantum spin systems, quantitative results for the generalized dilution problem
from large quantum Monte Carlo simulations remain inaccessible.

Angle resolved photo emission spectroscopic (ARPES) experiments
as well as ab-initio calculations
on a number of copper oxide materials 
provide strong evidence that an effective one-band Hubbard model description
of these systems must include direct hopping parameters $t'$ and
$t''$ to second and third nearest neighbor sites.
Furthermore, such experiments and calculations
indicate that these parameters are not significantly smaller than the
nearest neighbor hopping $t$, with $t'/t \sim -0.3$ and
$t''/t \sim 0.15$. 
We have recently included direct hopping parameters $t'$ and $t''$ 
in a derivation of a spin-only
Hamiltonian representation 
of the half-filled $t-t'-t''-U$ Hubbard model~\cite{Delannoy2}.
As a result of these sizeable energy scales,
our analysis of magnon excitation spectra
in La$_2$CuO$_4$ reveal that
the contributions from these parameters are of
similar magnitude to the four hop  (order $1/U^4$) processes for 
nearest neighbor hopping, which give rise to the ring exchange
interactions studied in Ref.~[\onlinecite{coldea01}] and in the present
paper (last term in $H_{\rm s}^{(4)}$ of Eq.~(\ref{Hs4})).

A key result  of Ref.~[\onlinecite{Delannoy2}]
is that the 
ground state staggered moment, approximately 0.235, is further reduced
from the value found for the $t-U$ Hubbard model, $\sim 0.27$,
using the $t$ and $U$ values of Coldea {\it et al.}~\cite{coldea01}.
This value is closer to, but undershoots
 the experimental estimate of $0.25$ ~\cite{Lee}.
Although this progression lies within the experimental uncertainty,
the detailed analysis of
Ref.~[\onlinecite{Delannoy2}]  suggests that the $t-t'-t''-U$ Hubbard model is a much
improved starting point for a quantitative description of the magnetic
properties of La$_2$CuO$_4$. This conclusion is in accordance with
ARPES studies and ab-initio calculations on various cuprates.

A natural extension of the work presented
in this paper would
be to investigate the role of $t'$ and $t''$ in the site dilution problem.
 In this model a large number of new ring
exchange terms are generated and the further neighbor hopping terms
allow for connected pathways for dilution concentrations above the
nearest neighbor percolation threshold.  It seems likely that these extra terms would 
change the shape of the $[M_{\rm s}(x)]$ vs $x$ curve,
specially for $x$ close to $x_p$, 
and hence change the qualitative aspect of the results even when the magnetization 
scale is factorized out of the problem, as  in Fig.~\ref{res15_fig}.

The realization that $t'$ and $t''$ are important energy scales in
a Hubbard model description of La$_2$CuO$_4$ leads to an interesting
experimental puzzle when considering the substitution of Cu$^{2+}$ by
non-magnetic Zn$^{2+}$ and/or Mg$^{2+}$.
As discussed earlier in this paper, and
as illustrated by the convergence of the
 the results of Fig.~\ref{res12_fig} for the Heisenberg model ($J_1$ only)
and the $t/U=1/8$ Hubbard model ($J_1$, $J_2$, $J_3$ and $J_c$), the
dilution-dependence of the electronic hopping pathways leads to
a crossover concentration $x^*$ ($x^*\sim 15\%$ for $t/U=1/8$) above
which the influence of the $J_2-J_3-J_c$ terms of order $1/U^3$
has essentially vanished.
However, the presence of direct $t'$ and $t''$ hoppings leads to 
additional {\it frustrating} second and third nearest neighbor
exchange with
 $J_2'\approx 4(t')^2/U$ and $J_3'' \approx 4(t'')^2/U$, respectively.
Unlike the $J_2$ and $J_3$ interactions generated by fourth order
hopping processes, $J_2'$ and $J_3''$ 
{\it do not depend} on the presence of
nearest neighbor pathways and hence are unaffected by the dilution
(see discussion in Section~\ref{No-charge-renorm}
and the one accompanying  Fig.~\ref{res9_fig}).
As there is now frustration which is independent 
of the existence of nearest neighbor pathways,
one would expect that upon
dilution there
would be a proliferation of Villain canted 
states~\cite{Villain-ZPhysB,Kinzel,Saslow1,Saslow2,Vannimenus,Gawiec1}
as the concentration of impurities approaches the
percolation threshold $x_p$. This could ultimately lead to  a 
Heisenberg spin glass phase for a dilution concentration $x<x_p$.
In this context, it is perhaps surprising that experiments
find sharp (resolution limited) magnetic Bragg peaks
in La$_2$Cu$_{1-x}$(Mg/Zn)$_x$O$_4$ all the way to $x=x_p$ \cite{Vajk}.
It would certainly be 
interesting to revisit this question and 
study in more detail the possibility of a spin glass phase
developing in 
La$_2$Cu$_{1-x}$(Mg/Zn)$_x$O$_4$ close to the percolation threshold.
We note further that in the region close to the percolation threshold
there is the possibility of a 
freezing transition of the transverse spin components only. Such a
transition could be observable in nuclear quadrupolar resonance
(NQR) or muon spin relaxation (muSR) experiments~\cite{Mirebeau-hyp}
as were done sometime ago
on  La$_2$Cu$_{1-x}$Zn$_x$O$_4$ ~\cite{Corti}. 
However, in those early experiments~\cite{Corti}, it 
now seems likely that the then detected transverse spin 
freezing was driven by doped holes introduced by an
imperfect control of the oxygen stoichiometry in 
 La$_2$Cu$_{1-x}$(Mg/Zn)$_x$O$_4$ ~\cite{Vajk,Vajk-review}.
It would be interesting to repeat such NQR and muSR experiments
on  La$_2$Cu$_{1-x}$(Mg/Zn)$_x$O$_4$ samples of the same quality
as those used in neutron scattering experiments of Ref.~[\onlinecite{Vajk}].


Another effect that could be relevant for
La$_2$Cu$_{1-x}$(Mg/Zn)$_x$O$_4$ is the local distortion
of the lattice due to the small difference in the ionic radius
between Cu$^{2+}$ and Mg$^{2+}$ or Zn$^{2+}$ ~\cite{Edagawa,Cherny-abs}.
This difference could lead to a local modification of the hopping
parameter $t$ in the neighborhood of a site where a Cu$^{2+}$ ion is
replaced by a nonmagnetic ion (see Fig. 2 in
Ref.~[\onlinecite{Edagawa}]).  Such disorder-induced variations of the
hopping parameters could then contribute to 
explain the difference between the experimental
data and QMC data in Fig.~\ref{science}.
The importance of local distortions could perhaps 
be provided by local probe experiments such as muSR, NMR or NQR.
This problem may also
be considered as a precursor to the study of
disorder-induced static magnetism in cuprate superconductors~\cite{Kampf}.
In this case the inclusion of mobile holes makes it
much more complicated, but the study of the 
diamagnetic dilution problem in 
La$_2$Cu$_{1-x}$(Mg/Zn)$_x$O$_4$
maintained at half
filling could provide a useful framework on which to build.


In conclusion, we have explored in this work the problem
of the evolution of the 
magnetic order in a spin-only representation of a 
site-diluted one-band Hubbard expressed in terms of a spin-only Hamiltonian, 
taking into account up to four hop processes.
For a finite ratio of hopping constant to on-site Coulomb
energy,  $t/U$, the resulting spin Hamiltonian differs from
the simpler site-diluted $S=1/2$ Heisenberg model, containing
effective exchange coupling beyond nearest neighbor as well
as ring exchange interactions. The long range exchange interactions,
the ring exchange and the renormalization of the nearest neighbor
exchange depend specifically on the local random hopping pathways
that remain uninterrupted by the missing (diluted) sites.
We hope that this study can motivate further analytical and numerical
studies of the site-diluted one-band Hubbard model as well as new
experiments on  La$_2$Cu$_{1-x}$(Mg/Zn)$_x$O$_4$ in the vicinity of
the percolation threshold.

\section{Acknowledgements}

It is a pleasure to thank A.-M. S. Tremblay for a related collaboration
leading to the publications of  Ref.~[\onlinecite{Delannoy,Delannoy2}]
as well as for useful comments on this manuscript.
We also thank
 G. Albinet,
 B. Castaing,
 A. Castro Neto,
 F. Delduc,
 T. Devereaux,
 A. Mucciolo,
 L. Raymond,
 A. Sandvik,
 R. Scalettar,
 O. Vajk and
 T. Vojta
for useful discussions. Partial support for this work was provided by NSERC
of Canada and the Canada Research Chair Program (Tier I) (M.G.) 
Research Corporation and the Province of Ontario (M.G.)
 and a Canada$-$France travel grant from the
French Embassy in Canada (M.G.and P.H.). M.G. acknowledges
the Canadian Institute for Advanced Research (CIFAR) for support.

\appendix

\section{Calculation procedure for the  Bogoliubov transformation \label{Calc}}

We summarize below the steps required to obtain the eigenvector
matrix  for $\tilde{H}$, $\widetilde{E}$ satisfying the boson
commutation relations (\ref{gfi}):

\begin{itemize}

\item Diagonalize $\tilde{H}$ using the Lapack routines. This
yields a set of eigenvalues $\omega_i$ with  corresponding
eigen-subspaces generated by the eigenvectors $E_i^n$, where
$n\in\{1,p_i\}$ and where $p_i$ is the degeneracy of the
eigenvalue and dimension of the subspace.

\item For the subspace $i$ define $E_i$, a $2N\times p_i$ matrix
of the corresponding eigenvectors $E_i^n$. Form the block matrix
$M_i$
\eqn{M_i\= E_i^{\dag} \tilde{I} E_i ,}
of size $p_i\times p_i$.

\item Invert $M_i$ to get $M_i^{-1}$.

\item Diagonalize $M_i^{-1}$, thus defining $K_i$ and $D_i$:
\eqn{M_i^{-1}\= K_i D_i K_i^{-1}.}

\item Define the matrix:
\eqn{A_i\=K_i\sqrt{\pm D_i}K_i^{-1}.}
In this expression, the sign $\pm$ corresponds to the sign of
$\omega_i$.

\item Define the new matrix of eigenvectors for the subspace $i$,
$\tilde{E_i}$:
\eqn{\tilde{E_i}\=E_i A_i.}

\item Repeat subspace by subspace to construct the eigenvector
matrix $\widetilde{E}$.

\end{itemize}

\section{Statistical errors}
\label{errors}

 this section we discuss the origin of the statistical errors we
find from our numerical results. Consider, as an example, the
lattice of size $L^2\=24\times 24$. For $x\=12\%$, we studied
${\cal N}_0 =620$ different realizations of disorder. The average
magnetization and root mean square (RMS) variation,
$ \Delta M$,  were found to be
\eqn{[ M(x=12\%,L=24)]_{\rm perc} \=
0.305\;, \Delta M= 0.0052}
from which we estimate the error on the measure to be $\pm
\sigma_{M(x)}$ ~\cite{Boas}
\begin{equation}
\sigma_{M(x)}= \f{\Delta M_{\rm s}}{{\sqrt{{{\cal N}_0}}}}.
\end{equation}
In this example the estimated error is thus extremely small,
around $0.1\%$ and the errors rise to around $1\%$ near the
percolation threshold. This small error estimate is consistent
with the statistical fluctuations observed in Figs.
\ref{res2_fig} and \ref{res4_fig}.

%
%
%

For the example considered above, 
the ratio of the dispersion to the mean value:
\eqn{\f{\Delta M}{[ M
]_{\rm perc}} \=  1.705 \%. } \label{delta_tot}
The ratio of the dispersion, $\Delta M$, to mean value
$[M]_{\rm perc}$, 
for fixed dilution $L$ as a function of $x$ 
and for fixed $x$ and as a function of $L$ 
are shown in Tables
(\ref{res1}) and (\ref{res2}), respectively.

%
%
%

\begin{table}[ht]
\begin{center}
\begin{tabular}{|c|D{.}{.}{3}|D{.}{.}{4}|D{.}{.}{3}|}
\hline
$x(\%)$ & \multicolumn{1}{l|}{$[ M]_{\rm perc}$} 
& \multicolumn{1}{l|}{$ \Delta M $} 
& \multicolumn{1}{l|}{$\f{ \Delta M }{[ M]_{\rm perc}} (\%)$} \\
\hline
0  &  0.338  &  0  &  0  \\
2  &  0.334  &  0.0014  &  0.419  \\
4  &  0.329  &  0.0022  &  0.668  \\
6  &  0.324  &  0.0028  &  0.864  \\
8  &  0.319  &  0.0036  &  1.128  \\
10  &  0.313  &  0.0041  &  1.310  \\
12  &  0.305  &  0.0052  &  1.705  \\
14  &  0.298  &  0.0060  &  2.013  \\
16  &  0.290  &  0.0070  &  2.414  \\
18  &  0.281  &  0.0080  &  2.847  \\
20  &  0.272  &  0.0096  &  3.529  \\
22  &  0.261  &  0.0109  &  4.176  \\
24  &  0.249  &  0.0135  &  5.421  \\
26  &  0.236  &  0.0154  &  6.525  \\
28  &  0.221  &  0.0183  &  8.281  \\
30  &  0.204  &  0.0213  &  10.440  \\
32  &  0.186  &  0.0237  &  12.742  \\
34  &  0.166  &  0.0274  &  16.506  \\
36  &  0.147  &  0.0291  &  19.795  \\
38  &  0.125  &  0.0316  &  25.280  \\
40  &  0.109  &  0.0330  &  30.275  \\
\hline
\end{tabular}
\caption{Staggered magnetization for $L^2=24\times 24$ and
$x\in[0,40]\%$}\label{res1}
\end{center}
\end{table}

\begin{table}
\begin{center}
\begin{tabular}{|c|c|c|}
\hline
$L$ & $[M]_{\rm perc}$ & $ \Delta M$\\
\hline
10  &  0.325  &  0.0181 \\
12  &  0.311  &  0.0166 \\
14  &  0.298  &  0.0160 \\
16  &  0.291  &  0.0134 \\
18  &  0.284  &  0.0118 \\
20  &  0.279  &  0.0114 \\
22  &  0.276  &  0.0103 \\
24  &  0.272  &  0.0096 \\
26  &  0.269  &  0.0090 \\
28  &  0.267  &  0.0079 \\
30  &  0.264  &  0.0080 \\
32  &  0.263  &  0.0075 \\
34  &  0.261  &  0.0068 \\
\hline
\end{tabular}
\caption{Staggered magnetization for $x=20\%$ and $L\in [10,34]$.
}\label{res2}
\end{center}
\end{table}
We can model this dispersion using three sources of variation:
firstly, for a given $x$ the number of magnetic sites varies from
configuration to configuration. Secondly, for fixed $N$ the number
of sites on the percolating cluster will also vary. Thirdly, there
will also be a contribution from configurational fluctuations for
a fixed number of sites. We stress that all these contributions
are quantum in origin. That is, the classical ground state is
perfectly ordered for all concentrations above the percolation
threshold, as discussed in the main text, hence at the classical
level, changing the number of sites, or local structures on the
percolating cluster will not change the order parameter. However,
the dilution reduces the local spin stiffness for spins in contact
with non-magnetic sites and increases the zero point spin
fluctuations. Hence these variations in number of sites and
structure change the value of the order parameter. Indeed this
point is already manifest by the fact that $[M
]_{\rm perc}$ decreases with $x$.

If $N_i(L^2,x)$ is the number of sites for realization $i$, and
the mean number of sites is defined in Eq. (\ref{moyennage})
then for the example considered we find
\eqn{\bar{N}(L=24,x=12\%) \=  506.8 \pm 7.7 .}
Hence
 \eqn{\f{\Delta\bar{N}}{\bar{N}} \= 1.51 \%.}
To check the importance of fluctuations in the number of
participating sites on the percolating cluster, $(N_{\rm perc})_i$ we
analyze the ratio $\f{(N_{\rm perc})_i}{N_i}$. We find
\eqn{\f{N_{\rm perc}}{N} \= 0.999 \pm 6.67\times 10^{-3}~,}
which gives
\eqn{\f{\Delta \f{N_{\rm perc}}{N} }{ \f{N_{\rm perc}}{N}} \= 0.6\%~.}
For a fixed number of magnetic  sites, we can define the quantity
\[\left(\f{\Delta M}{[ M
]_{\rm perc}}\right)_{\rm mag}\]
as a measure of the configurational contribution to the dispersion
in ground state order parameter values, where
$\left(\dots\right)_{\rm mag}$ is the disorder average over the
restricted set of configurations with $N_i=constant$. For the example
discussed here we find
\eqn{\left(\f{\Delta M}{[ M
]_{\rm perc}}\right)_{\rm mag} \= 0.4677\%~,}
from which we estimate the total dispersion
\eqn{\left(\f{\Delta M}{ [M]_{\rm perc}}\right)_{\rm tot} \=
\f{\Delta\bar{N}}{\bar{N}}  +  \f{\Delta \f{N_{\rm perc}}{N} }{
\f{N_{\rm perc}}{N}} + \left(\f{\Delta M}{[
M]_{\rm perc}}\right)_{\rm mag}   \simeq 2\%~.}
in good agreement with Eq.~(\ref{delta_tot}). The analysis can be
generalized to the other values of $x$ and $L$. For example for
$L^2=24\times 24$ and $x\=30\%$ we have:
\eqn{\left\{\begin{array}{ccl}
\f{\Delta\bar{N}}{\bar{N}} &=& 3.48 \%~,\\
\\
\f{\Delta \f{N_{\rm perc}}{N} }{ \f{N_{\rm perc}}{N}} &=& 1.41\%~,\\
\\
\left(\f{\Delta M}{[ M
]_{\rm perc}}\right)_{\rm mag} &=& 5.45\%~,
\end{array}\right.}
which correspond to $\left(\f{\Delta M}{ [M]_{\rm perc}}\right)_{\rm tot} \simeq
10\%$, as obtained in Table \ref{res1}. Hence this analysis seems
to account for the dispersion in magnetization values to a good
level of approximation. The three sources of dispersion are of the
same order of magnitude as long as one remains well away from the
percolation threshold. At low defect concentration (small $x$) it
is the fluctuations in the number of magnetic sites that
dominates. As $x$ increases the fluctuations increase, as one
might expect as one approaches the critical percolating regime,
and at large $x$ it is the configurational contribution for fixed
particle number which dominates.  At $40\%$ dilution the
dispersion in values approaches $30\%$ of the mean order parameter
value. Despite this large dispersion for this value of $x$ the
number of configurations, ${\cal N}_0=2000$, is large enough to
keep the estimated error at the $1\%$ level.

\bibliographystyle{apsrev}
\bibliography{bibli_michel}

\begin{thebibliography}{67}
\bibitem[{Not()}]{Note-ttt}
\bibinfo{note}{A recent study~\cite{Delannoy2} that incorporates the efect of
  second ($t'$) and third ($t''$) nearest neighbor oppings also find that a
  value $t/U \sim 1/8$ allows for a spin-only Hamiltonian whose magnon
  excitations suitably describe the spin-waves of La$_2$CuO$_4$
  ~\cite{coldea01}.}
\end{thebibliography}

\end{document}